\documentclass[12pt, draftclsnofoot, onecolumn]{IEEEtran}
\newcommand{\twocols}{false}

\newlength{\figurewidth}
\usepackage{ifthen}

\usepackage[pdftex]{graphicx}
\graphicspath{{./figures/}}
\usepackage[usenames,dvipsnames]{xcolor}

\usepackage{amsmath,amssymb,amsthm,amsfonts}
\usepackage{nicefrac}
\usepackage{cite}
\usepackage{color}
\usepackage{transparent}

\ifCLASSOPTIONcompsoc
\usepackage[caption=false,font=normalsize,labelfont=sf,textfont=sf]{subfig}
\else
\usepackage[caption=false,font=footnotesize]{subfig}
\fi

\newcounter{tempEquationCounter}
\newcounter{thisEquationNumber}

\allowdisplaybreaks

\DeclareMathOperator*{\argmin}{arg\,min}
\newcommand{\defined}{{\;\overset{\Delta}{=}\;}}

\newcommand{\rate}{R}

\newcommand{\AWGN}{\mathrm{AWGN}}
\newcommand{\fracscheduling}{a}

\newcommand{\nt}{n_\mathrm{t}}
\newcommand{\nd}{n_\mathrm{d}}
\newcommand{\ndl}{n_\mathrm{t,DL}}
\newcommand{\ntot}{n_\mathrm{tot}}
\newcommand{\Rate}{R}

\newcommand{\perror}{\varepsilon}
\newcommand{\pout}{\varepsilon_\mathrm{out}}

\newcommand{\fblequiv}{\mathrm{b}}

\newcommand{\Ub}{{U_{\mathrm{b}}}}

\newcommand{\Ktot}{K_\mathrm{tot}}
\newcommand{\Kavg}{\overline{K}}
\newcommand{\K}{K}
\newcommand{\Nt}{M}
\newcommand{\suf}{T}
\newcommand{\sufss}{^{(\suf)}}
\newcommand{\rateadapt}{\Phi}
\newcommand{\rateadaptopt}{\Phi^*}
\newcommand{\Nsnrpoints}{N_\musig}
\newcommand{\Nrates}{N_\rate}

\newcommand{\Ptot}{P_\Sigma}
\newcommand{\Set}{\mathcal{K}}
\newcommand{\wtotal}{w}
\newcommand{\wnormal}{w}

\newcommand{\power}{\rho}

\newcommand{\fixpower}{\rho}

\newcommand{\aslot}{\mathsf{A}}
\newcommand{\bslot}{\mathsf{B}}

\newcommand{\expected}[1]{\mathbb{E}\left[#1\right]}
\newcommand{\real}[1]{\Re\left\{#1\right\}}

\newcommand{\Prob}[1]{\mathbb{P}\left\{#1\right\}}

\newcommand{\LL}{\mu}

\newcommand{\svec}{\mathbf{s}}

\newcommand{\xvec}{\mathbf{x}}

\newcommand{\yvec}{\mathbf{y}}

\newcommand{\zvec}{\mathbf{z}}
\newcommand{\hvec}{\mathbf{h}}
\newcommand{\hmmse}{\hat{\mathbf{h}}}
\newcommand{\esterror}{\mathbf{e}}
\newcommand{\Hmat}{\mathbf{H}}
\newcommand{\Hmmse}{\hat{\mathbf{H}}}
\newcommand{\Pmat}{\mathbf{P}}
\newcommand{\vvec}{\mathbf{v}}
\newcommand{\Vmat}{\mathbf{V}}
\newcommand{\Ximat}{\mathbf{\Xi}}
\newcommand{\MMPI}{\dag}
\newcommand{\herm}{\mathsf{H}}
\newcommand{\eye}{\mathbf{I}}
\newcommand{\trace}[1]{\mathrm{tr}\left(#1\right)}
\newcommand{\sinr}{\mathsf{SINR}}
\newcommand{\Intsum}{{I_\Sigma}}
\newcommand{\shape}{\nu}
\newcommand{\gammao}{\gamma_\mathrm{o}}
\newcommand{\Sigpower}{G}

\newcommand{\Sigpowergauss}{{\tilde{G}}}
\newcommand{\sigpower}{g}
\newcommand{\csifbl}{\mathrm{IC,F}}
\newcommand{\Gcsifbl}{\tilde{G}_\csifbl}

\newcommand{\sigman}{\sigma_\mathrm{e}}
\newcommand{\sigmansq}{\sigma_\mathrm{e}^2}
\newcommand{\musig}{{\hat{g}}}
\newcommand{\mutilde}{{\tilde{\mu}}}
\newcommand{\binl}{{l}}
\newcommand{\lambdac}{{\lambda_\mathrm{c}}}
\newcommand{\sigmasig}{{\sigma_\Sigpower}}
\newcommand{\sigmasigsq}{{\sigma_{\Sigpower}^2}}
\newcommand{\Puplink}{P_\mathrm{UL}}
\newcommand{\dispersion}{\mathcal{V}}

\newcommand{\Mellin}{\mathcal{M}}

\newcommand{\s}{{\theta}}
\newcommand{\salt}{\tilde{\theta}}

\newcommand{\ta}{{t_1}}
\newcommand{\tb}{{t_2}}
\newcommand{\ts}{{t}}

\newcommand{\sts}{}
\newcommand{\stsb}{}
\newcommand{\tsk}{}

\newcommand{\Mfun}[1]{\mathbb{K}\left(#1\right)}

\newcommand{\Mfunsuper}[1]{\mathbb{K}^{(T)}\left(#1\right)}

\newcommand{\pv}{{p_\mathrm{v}}}
\newcommand{\pvk}{{p_{\mathrm{v},k}}}

\newcommand{\Abit}{\mathit{A}}

\newcommand{\Sbit}{\mathit{S}}
\newcommand{\Abitcum}{{\mathbf{A}}}
\newcommand{\Dbitcum}{{\mathbf{D}}}
\newcommand{\Sbitcum}{{\mathbf{S}}}

\newcommand{\Delay}{{\mathit{W}}}
\newcommand{\Asnr}{\mathcal{A}}

\newcommand{\Ssnr}{\mathcal{S}}

\newcommand{\Xsnr}{\mathcal{X}}

\IEEEoverridecommandlockouts 
\begin{document}
\title{Delay Performance of the Multiuser MISO Downlink under Imperfect CSI and Finite Length Coding}

\author{
  \IEEEauthorblockN{Sebastian~Schiessl\IEEEauthorrefmark{1}, James~Gross\IEEEauthorrefmark{1}, Mikael~Skoglund\IEEEauthorrefmark{1} and Giuseppe~Caire\IEEEauthorrefmark{3}}\\
  \IEEEauthorblockA{\IEEEauthorrefmark{1}School of Electrical Engineering and Computer Science, KTH Royal Institute of Technology, Stockholm, Sweden}\\
  \IEEEauthorblockA{\IEEEauthorrefmark{3} Institute for Telecommunication Systems, Technical University Berlin, Germany}\\
  Emails: $\{$schiessl,jamesgr,skoglund$\}$@kth.se, and caire@tu-berlin.de
}
\date{\today}
\maketitle

\newtheorem{lemma}{Lemma}
\newtheorem{assumption}{Assumption}
\newtheorem{corollary}{Corollary}
\newtheorem{result}{Result}


\begin{abstract}
We use stochastic network calculus to investigate the delay performance of a multiuser MISO system with zero-forcing beamforming. First, we consider ideal assumptions with long codewords and perfect CSI at the transmitter, where we observe a strong channel hardening effect that results in very high reliability with respect to the maximum delay of the application. We then study the system under more realistic assumptions with imperfect CSI and finite blocklength channel coding. These effects lead to interference and to transmission errors, and we derive closed-form lower and upper bounds on the resulting error probability. Compared to the ideal case, imperfect CSI and finite length coding cause massive degradations in the average transmission rate. Surprisingly, the system nevertheless maintains the same qualitative behavior as in the ideal case: as long as the average transmission rate is higher than the arrival rate, the system can still achieve very high reliability with respect to the maximum delay.
\end{abstract}


\section{Introduction}
\label{sec:introduction}
Due to the random nature of the wireless channel, it is notoriously difficult to design wireless communication systems for applications that require both very low latency and very high reliability.
For example, applications in factory automation often require latencies of just a few milliseconds and reliability (with respect to this deadline) of $1-10^{-8}$ and above~\cite{osseiran2014scenarios,3gpp.22.804}, which is difficult to achieve in a wireless channel that is subject to fading and noise.
In order to increase the reliability of the system, one can equip the transmitter with multiple antennas, which is known as multiple-input single-output (MISO).
When transmitting only to a single receiver, multiple transmit antennas increase the diversity of the system and reduce the variations in the signal strength in fading channels, leading to more reliable transmissions. When the transmitter has channel state information (CSI), it can use beamforming to send the signal in the direction of the user's channel, resulting not only in a diversity gain but also a power gain~\cite{tse2005fundamentals}, making the system even more resilient against errors.
On the other hand, a transmitter with $\Nt$ antennas can also serve $K\le \Nt$ different users at the same time, i.e., achieve a multiplexing gain. Serving multiple users at once means that each user can be scheduled more often, which can reduce the delay. An often used transmission strategy for the multiuser MISO downlink is zero-forcing beamforming (ZFBF), which ensures that the signal intended for each user does not create interference at the other (unintended) receivers.
Nevertheless, increasing $\K$ reduces the beamforming gain, i.e., reduces the data rates of the individual transmissions.
The trade-off between the multiplexing gain and the beamforming gain was studied in \cite{hochwald2002space} with respect to the ergodic capacity. 

However, the ergodic capacity does not accurately reflect the delay performance of the system. When there is a certain probability that the data rate is small, or when transmission errors occur, the transmitter must keep the data in a buffer so that it can be transmitted in subsequent time slots. This buffering causes a random delay. The queueing delay may sometimes grow until the deadline of the application is violated. For applications that require very high reliability, the communication system must be designed such that the probability of a deadline violation is minimized. For example, low beamforming gains should be avoided, as low beamforming gains increase the probability that the individual data rates are small. In \cite{schiessl2018misoglobecom}, we studied the trade-off between the multiplexing and beamforming gains with respect to the queueing performance.

Fortunately, as the number of antennas $\Nt$ grows, the transmitter can schedule $K=\fracscheduling \Nt$ users with $a<1$ and thus benefit from a linear increase in both multiplexing gain and beamforming gain. As the beamforming gain increases linearly in $\Nt$, the system will experience only small variations (relative to the average) in the achievable transmission rate. This effect is known as channel hardening \cite{hochwald2004hardening}. In this case, we suspect that only the \emph{average} of the transmission rate will determine the queueing performance: When the \emph{average} transmission rate is higher than the incoming data rate at the transmit buffer, then long queueing delays should be very unlikely (occur with almost zero probability); otherwise, long queueing delays have probability one.

In this context, it is of critical importance that the reliability of the physical layer transmissions is modeled accurately. Specifically, when the duration of each time slot is short, channel estimation causes a significant overhead, and the transmitter can only acquire an imperfect estimate of the channel state. First of all, this means that zero-forcing beamforming cannot eliminate the interference. Second, the transmitter does not know the actual signal-to-interference-and-noise ratio (SINR) of the channel, and therefore, outages can occur when the actual channel capacity is below the selected transmission rate. The transmitter must then find a careful balance between the outage probability and the rate with respect to the queueing delay.
In addition to imperfect CSI, we note that the transmitter cannot achieve error-free transmissions at the channel capacity when the blocklength of the channel code is finite. All of these effects must be taken into account when considering systems for ultra low latency communications.

\subsection{Related Work}
This paper builds on results from several research areas. On the physical layer, we consider the multiuser MIMO downlink with imperfect CSI, as well as finite blocklength coding. On top of these physical-layer aspects, we investigate the queueing delay of the data at the link layer.

\subsubsection{Multiuser-MIMO and Imperfect CSI} Linear ZFBF precoding in the multiuser MISO downlink has been studied by several authors.
Although ZFBF is not capacity-achieving, Yoo and Goldsmith \cite{yoo2006optimality} showed that when the total number of users $\Ktot$ is much larger than the number of antennas $\Nt$, then ZFBF can achieve the same asymptotic performance as the capacity-achieving scheme based on dirty-paper coding (DPC) \cite{caire2003achievable}. 
These results hold only if the transmitter has channel state information (CSI) of all users. While the authors investigated in \cite{yoo2007multi} also the impact of quantized channel state feedback, it would still be impossible to receive feedback from e.g. 100 or more users when the duration of each time slot is short. 
The random beamforming scheme by Sharif and Hassibi \cite{sharif2005capacity} reduces the overhead from collecting CSI by transmitting a training sequence along a set of randomly created beams. The users then send the index and the signal-to-interference-and-noise ratio (SINR) of the beam with the highest SINR. However, in this scheme, some of the users may not be scheduled for a long period because the scheduling decision is based on the random SINR. Furthermore, even though the overhead from collecting CSI is reduced, collecting feedback from many users may still be infeasible when considering scenarios with very low latency.
In fact, Ravindran and Jindal \cite{ravindran2012multiuser6247447} found that, given a fixed budget for the total overhead, it is better to collect accurate channel estimates from only a small number of users than to collect inaccurate CSI from many users, i.e., they found that accurate CSI is more important than multiuser diversity.
When the transmitter has only CSI for the users that are scheduled, Zhang et al. \cite{zhang2009mode} studied whether the transmitter should send data to a single user or to $K=\Nt$ users.
The same authors studied in \cite{zhang2011multi} the more general case $K=\fracscheduling\Nt$ and also considered imperfect CSI. This is very close to our work, but the authors studied only the ergodic sum rate and assumed that an additional perfect feedback link provided the exact value of the channel capacity to the transmitter, such that outages did not occur. Similarly, the authors in \cite{caire2010multiuser} studied the ergodic capacity of a multiuser MISO system under imperfect CSI. In the ergodic case, rate adaptation is not necessary, and the performance loss is due to imperfect beamforming and due to additional noise terms at the receiver. 
In contrast, we want to study the queueing performance of a system where outages occur because the transmitter must adapt the rate to an imperfect estimate of the channel, and where the receiver must decode the signal in the same time slot (as opposed to decoding over an infinite time horizon).

\subsubsection{Finite Blocklength Coding}
Some well-known results on channel coding at finite blocklength were derived by Polyanskiy et al. \cite{polyanskiy2010channel}, who showed that the loss in the achievable data rate due to finite blocklength can be approximated by a simple second-order expression. Yang et al. extended these results to fading channels \cite{yang2014quasi}. These works generally assume non-Gaussian codebooks in Gaussian noise. 
Therefore, the results cannot be applied when the transmissions create mutual interference. 
Scarlett et al.~\cite{scarlett2017dispersion} studied the performance of Gaussian codebooks under non-Gaussian interference. However, we are not aware of any results combining finite blocklength coding with multiuser MIMO.

\subsubsection{Delay Analysis} 
The queueing delay, which occurs due to transmission errors and low transmission rates, can for example be analyzed through the frameworks of stochastic network calculus \cite{fidler2006network,alzubaidy2016ton} or effective capacity \cite{wu2003effective}. In our previous work \cite{schiessl2016imperfectcsi}, we applied stochastic network calculus to a single-antenna channel with imperfect CSI and finite length coding.
With respect to MIMO systems, the effective capacity of the single-user MIMO channel was studied in several works \cite{liu2008effective4595458,jorswieck2010effective5550915,gursoy2011mimo6006584}.  
The multi-user case was studied by Li et al. \cite{li2013adaptive}, but this work does not fit our assumptions because the authors assumed that the channels are non-fading.

\subsection{Contributions}
In this work, we study the multiuser MISO downlink both under ideal assumptions and under more realistic assumptions with imperfect CSI and finite blocklength coding. Specifically, we make the following contributions:

\begin{itemize}
	\item For the ideal scenario with perfect CSI and long codewords, we use our previous results \cite{schiessl2018misoglobecom} to study the effect of channel hardening. We investigate how many antennas are necessary to achieve extremely high reliability with almost zero violations of the deadline.
	\item For the realistic scenario with imperfect CSI, we derive two closed-form approximations, corresponding to lower and upper bounds on the conditional outage probability. 
	\item For the realistic scenario with imperfect CSI and finite blocklength of the channel code, we derive a closed-form approximation for the conditional error probability.
	\item We verify by extensive Monte Carlo simulations that the derived expressions are lower or upper bounds on the conditional outage or error probability.
	\item We show that the closed-form expressions can be used to find a rate adaptation function that minimizes the delay violation probability (based on network calculus).
	\item Our numerical analysis shows that imperfect CSI leads to substantial losses in the average transmission rate. Furthermore, we find that the additional loss due to finite blocklength effects is only relevant when the CSI becomes nearly perfect. Despite the massive performance loss compared to the ideal case, a system that shows a strong channel hardening effect in the ideal scenario maintains this behavior qualitatively in the realistic case. In other words, multiuser MISO systems can achieve very high reliability even under non-ideal assumptions.
\end{itemize}

This paper is structured as follows: In Sec.~\ref{sec:system_model}, we present the system model for the ideal scenario with perfect CSI and long codewords. In Sec.~\ref{sec:analysis}, we present a short summary of the delay analysis through network calculus, and perform the delay analysis for the ideal scenario.
In Sec.~\ref{sec:analysis_icsi}, we show how imperfect CSI and finite blocklength effects can be modeled and analyzed.
Numerical evaluations are presented in Sec.~\ref{sec:numerics}, before we finally conclude the paper in Sec.~\ref{sec:conclusions}.

\section{System Model}
\label{sec:system_model}
We consider a system where data is sent from a transmitter with~$\Nt$ antennas to $\Ktot$ single-antenna users, with $\Ktot \gg \Nt$. We assume time-slotted transmissions. In each time slot, the transmitter can select only a subset $\Set\subset \{1,\ldots,\Ktot\}$ of users, with the number of scheduled users denoted as $K\defined|\Set|\leq \Nt$. As we consider scenarios where the duration of each time slot is short, the overhead from collecting channel state information (CSI) for all $\Ktot$ users would be overwhelming. Thus, the transmitter cannot select the user set $\Set$ based on the instantaneous CSI. Instead, the transmitter selects in each time slot a set of users $\Set$ and then collects channel state information only for those users, similar to \cite{zhang2011multi}.
In this section, we describe a basic model with ideal assumptions based on \cite{schiessl2018misoglobecom}, i.e., we assume that the transmitter has perfect CSI for the scheduled users and that data can be transmitted at a rate equal to the channel capacity without errors. 
The more realistic scenario with imperfect CSI and finite blocklength of the channel code will be modeled and analyzed in Sec.~\ref{sec:analysis_icsi}.

First, we describe in Sec.~\ref{ssec:system_data_phy} the data transmission from a physical layer perspective. In Sec.~\ref{ssec:link_layer_strategies}, we discuss user scheduling. In Sec.~\ref{ssec:system_link}, the system is described from a queueing perspective. Finally, we present the problem statement in Sec.~\ref{ssec:problem_statement}.

\subsection{Physical Layer Model}
\label{ssec:system_data_phy}
The received signal $\yvec\sts\in\mathbb{C}^{K\sts\times 1}$ at the $K\sts$ scheduled users can be described as
\begin{equation}
\yvec\sts = \Hmat^\herm\sts \xvec\sts +\zvec\sts
\,.
\end{equation}
For the channel matrix $\Hmat\sts=[\hvec_1,\ldots,\hvec_K]\in\mathbb{C}^{\Nt \times K}$, we assume Rayleigh fading, i.e., all elements are independent and identically distributed (i.i.d.) with Gaussian distribution $\mathcal{CN}(0,1)$. Furthermore, we consider the quasi-static fading model where the channel $\Hmat$ remains constant for the duration of one time slot, consisting of $\nd$ channel uses, and changes to an independent realization in the next time slot (note that the set $\Set$ of scheduled users also changes). The input signal is denoted as $\xvec\sts\in\mathbb{C}^{\Nt \times 1}$ and must satisfy a short-term power constraint $\trace{\expected{\xvec\sts\xvec\sts^\herm}}\le \Ptot$ for each realization of $\Hmat\sts$. The noise $\zvec\sts \in\mathbb{C}^{K\sts\times 1}$ has i.i.d. components $\mathcal{CN}(0,1)$. 

The transmitter encodes the data for the $K\sts$ scheduled users into code symbols $\xvec\sts\in\mathbb{C}^{\Nt \times 1}$ (one symbol per antenna). In order to obtain $\xvec$, the transmitter can encode the data of the $K$ users individually into symbols $\svec\sts\in\mathbb{C}^{K\times 1}$ and apply a precoding strategy to obtain $\xvec$ from $\svec$. We focus in this work on Zero-Forcing Beamforming (ZFBF), which is a linear strategy that completely eliminates the interference of the signals at the other receivers.
In this case, the input signal vector $\xvec\stsb$ is given by \cite{caire2003achievable,schiessl2018misoglobecom}
\begin{equation}
	\xvec\stsb =\Vmat\stsb\Pmat\stsb^{1/2}\svec\stsb
\end{equation}
where $\Vmat\stsb=[\vvec_1,\ldots,\vvec_K]$ is the precoding matrix and $\Pmat\stsb=\mathrm{diag}(\power_{\tsk 1},\ldots,\power_{\tsk K})$ is the power allocation matrix. We require that the sum power $\Ptot$ is allocated equally to all users, i.e., $\power_{1}=\ldots=\power_{K}=\Ptot/K$. The vector $\svec\stsb$ denotes the $K\stsb\times 1$ vector of (independently) coded Gaussian symbols for the $K\stsb$ scheduled users.
When the transmitter perfectly knows the channel matrix $\Hmat$, the ZFBF precoder is given as \cite{caire2003achievable,schiessl2018misoglobecom}
\begin{equation}
	\Vmat\stsb=\Hmat\stsb^\MMPI \Ximat\stsb^{1/2}
\end{equation}
where $\Hmat\stsb^\MMPI = \Hmat\stsb(\Hmat^\herm\stsb\Hmat\stsb)^{-1}$ is the Moore-Penrose pseudo-inverse of $\Hmat^\herm\stsb$ and $\Ximat\stsb=\mathrm{diag}\left(\xi_{\tsk 1},\ldots,\xi_{\tsk K}\right)$ is the normalization matrix such that the columns of $\Vmat\stsb$ have unit-2 norm. The variables $\xi_{\tsk k}$ are central chi-square distributed (scaled by a factor $1/2$) with $2 m\stsb$ degrees of freedom, where $m\stsb = \Nt-K\stsb+1$. Their PDF is given by \cite[Lemma~4]{caire2003achievable}
\begin{equation}
	f_m(\xi)=\frac{1}{\Gamma(m)} \xi^{m-1} e^{-\xi}
	\label{eq:pdf_central_chisquare_scaled}
	\;.
\end{equation}
We assume for now that the blocklength $\nd$ of the channel code is sufficiently long, so that the system can achieve error-free transmission ($\perror=0$) to user $k$ at a rate \cite{caire2003achievable,schiessl2018misoglobecom}
\begin{equation}
	R_{k}=
	\log_2(1+\power_{\tsk k}|\hvec_k\vvec_k|^2)=
	\log_2(1+\power_{\tsk k}\xi_{\tsk k})
	\,,
	\label{eq:rates_zfbf}
\end{equation}
which changes along with $\Hmat$ from time slot to time slot.

\subsection{Scheduling}
\label{ssec:link_layer_strategies}
In each time slot, the transmitter can schedule only a subset $\Set$ of users. To make sure that each user is scheduled regularly, we consider superframes of length $\suf$ slots, and we require that each user is scheduled exactly once within a superframe. 
The average number of scheduled users per slot is given as $\Kavg=\Ktot/ \suf$. However, $\Kavg$ may not always be integer.
To simplify notations and discussions, our analysis considers only the case where $\Kavg$ is integer, i.e., where the transmitter schedules a constant number of $K=\Kavg$ users in each time slot. The analysis of non-integer $\Kavg$ is discussed in Appendix~\ref{appendix_groupsAB}.

\subsection{Link Layer Model}
\label{ssec:system_link}
In time slot $\ts$, $A_k(\ts)$ data bits intended for downlink transmission to user $k$ arrive at the transmitter. The data is stored in a transmit buffer, with individual buffers (or queues) for each user.
We assume that the arrival process $A_k(\ts)$ is constant over time and equal for all users, with $\alpha$ denoting the constant number of bits that arrive at the queue of each user in each time slot.
In the first part of this work, we assume error-free transmissions, and the service rate offered by the wireless system in each time slot to user $k$ is given as $S_k(\ts)=\nd R_{k}(\ts)$ when $k$ is among the scheduled users, or $S_k(\ts)=0$ when user $k$ is not scheduled. When transmission errors occur with probability $\perror>0$, a scheduled user $k$ is served with $S_k(\ts)=\nd R_{k}(\ts) Z$, where $Z\sim\mathrm{Bernoulli}(1-\perror)$.
The departure process $D_k(\ts)$ describes the amount of data that is transmitted to the receiver. Thus, $D_k(\ts)$ is limited both by the amount of data waiting in the buffer, as well as by the service rate $S_k(\ts)$.
The cumulative arrival, service, and departure processes are defined as
\begin{equation}
\Abitcum_k(\ta,\tb) \defined \sum\limits_{\ts=\ta}^{\tb-1}A_k(\ts) \;,
\quad\Sbitcum_k(\ta,\tb) \defined \sum\limits_{\ts=\ta}^{\tb-1}S_k(\ts) \;,
\end{equation}
\begin{equation}
\quad\Dbitcum_k(\ta,\tb) \defined \sum\limits_{\ts=\ta}^{\tb-1}D_k(\ts) \;.
\end{equation}
The queueing delay $\Delay_k(\ts)$ of user $k$ at time $\ts$ is defined as the time it takes for all data that arrived prior to time $\ts$ to depart from the transmit buffer and reach the receiver \cite{alzubaidy2016ton, schiessl2016imperfectcsi}:
\begin{equation}
\Delay_k(\ts) \defined \inf\left\{u\geq 0:\quad \Abitcum_k(0,\ts) \leq \Dbitcum_k(0,\ts+u) \right\} \;.
\label{eq:def_delay}
\end{equation}
The delay $\Delay_k(\ts)$ is random. The reliability of a communication system with respect to the deadline $\wtotal$ of the application can be described by the probability that the random delay $\Delay_k(\ts)$ of the data for user $k$ exceeds the target delay $\wtotal$ at any time $\ts$:
\begin{equation}
\pvk(\wtotal) \defined \sup_{\ts\ge 0} \left\{ \Prob{\Delay_k(\ts)>\wtotal} \right\}\; .
\label{eq:def_pdelayviol_time}
\end{equation}
We note here that for the considered system, the delay violation probability $\pvk(\wtotal)$ cannot be analyzed through closed-form expressions.
We thus follow our previous works \cite{schiessl2016imperfectcsi,schiessl2018misoglobecom}
and use stochastic network calculus \cite{fidler2006network,alzubaidy2016ton} to obtain analytical bounds on $\pvk(w)$.

\subsection{Problem Statement}
\label{ssec:problem_statement}
In the first part of this work, we consider a system with perfect CSI and long blocklength of the channel code. In this case, error-free transmissions at the channel capacity can be achieved. In our previous work \cite{schiessl2018misoglobecom}, we analyzed the optimal number of scheduled users $\Kavg$ such that the delay violation probability $\pvk(w)$ is minimized, i.e., the optimal trade-off between the multiplexing gain and the beamforming gain.
In this work, we study the effect of channel hardening:
when the number of antennas $\Nt$ grows, the data rate of the wireless channel becomes nearly constant. Due to channel hardening, we expect that the system will become very reliable, i.e., that long queueing delays occur with very low (almost zero) probability when the average transmission rate is above the arrival rate. Naturally, when the average transmission rate is below the arrival rate, the queueing delay grows to infinity and the delay violation probability is one. We will investigate how many antennas $\Nt$ are necessary to observe such a zero/one behavior in practice.

In the second part of this work, we consider the same question in a more realistic scenario, where the transmitter must first estimate the channel before the transmission starts, and where the blocklength of the channel code is finite. As we will discuss in Sec.~\ref{sec:analysis_icsi}, imperfect CSI and finite blocklength coding may have a significant impact on the system performance in the realistic case. Most importantly, scheduling a larger number of users $\Kavg$ will increase the interference and also result in a larger overhead for channel estimation. We therefore expect that the optimal number of scheduled users $\Kavg$ will decrease. However, it is not clear whether these effects will just lead to a change in the optimal value of $\Kavg$ and to a \emph{quantitative} loss in the overall performance, or whether these effects lead to a \emph{qualitative} change in the system performance. Specifically, we want to find out whether a realistic system maintains the zero/one behavior with respect to the delay distribution, i.e., whether the system still shows extremely high reliability whenever the average transmission rate is above the arrival rate.

\section{Analysis -- Ideal Case}
\label{sec:analysis}
In this section, we follow \cite{schiessl2018misoglobecom} and outline the analytical approach to determine $\pvk(w)$. Specifically, in Sec.~\ref{ssec:queueing}, we present a summary of the delay analysis through stochastic network calculus in a transform domain \cite{alzubaidy2016ton}. 
In Sec.~\ref{ssec:queueing_rr} we show how these results can be used when the users are only scheduled once per superframe. In Sec.~\ref{ssec:delay_analysis}, we analytically obtain the stochastic network calculus bounds for the ideal case. 
We assume that all users are subject to the same channel characteristics and delay requirements, and we drop the subscript $k$ to shorten the notation.

\subsection{Stochastic Network Calculus (SNC)}
\label{ssec:queueing}
This section closely follows our previous works \cite{schiessl2016imperfectcsi,schiessl2018misoglobecom} and provides a summary of stochastic network calculus \cite{fidler2006network,alzubaidy2016ton}.
 
The delay $\Delay(t)$ in (\ref{eq:def_delay}) is defined in terms of the arrival and departure processes. 
However, the distribution of the delay can be found directly from the statistics of the arrival and service processes.
We follow \cite{alzubaidy2016ton} and describe these processes in the exponential domain, also referred to as \emph{SNR domain}. The arrival and service processes in the bit domain, $\Abit(\ts)$ and $\Sbit(\ts)$, are converted to the SNR domain (denoted by calligraphic letters) as 
\begin{equation}
\Asnr(\ts) \defined e^{\Abit(\ts)}\,,\quad\Ssnr(\ts) \defined e^{\Sbit(\ts)}
\,.
\end{equation}
We assume constant arrivals with $\Abit(\ts)=\alpha$.
Consider for now a service process $\Sbit(\ts)$ that is independent and identically distributed (i.i.d.) between time slots. 
Then, an upper bound on the delay violation probability $\pv(w)$ can be obtained in terms of the Mellin transforms of $\Asnr$ and $\Ssnr$.
The Mellin transform $\Mellin_\Xsnr(\s)$ of a nonnegative random variable $\Xsnr$ is defined as \cite{alzubaidy2016ton}
\begin{equation}
\Mellin_\Xsnr(\s)\defined\mathbb{E}\left[\Xsnr^{\s-1}\right]
\end{equation}
for a parameter $\s \in \mathbb{R}$. 
For the analysis, we choose $\s>0$ and define the kernel \cite{alzubaidy2016ton,schiessl2015delay}
\begin{align}
\Mfun{\s,\wnormal} &\defined \lim\limits_{t\to\infty} \sum_{u=0}^{t} \Mellin_{\Asnr}(1+\s)^{t-u} \cdot \Mellin_{\Ssnr}(1-\s)^{t+\wnormal-u}
\,.
\end{align}
Under the condition $\Mellin_{\Asnr}(1+\s)\Mellin_{\Ssnr}(1-\s) < 1$, the kernel converges:
\begin{align}
\Mfun{\s,\wnormal}&= \frac{\Mellin_{\Ssnr}(1-\s)^{\wnormal}}{1-\Mellin_{\Asnr}(1+\s)\Mellin_{\Ssnr}(1-\s)} 
\;.
\label{eq:snc_kernel}
\end{align}
For any parameter $\s>0$, the kernel $\Mfun{\s,\wnormal}$ provides an upper bound on the delay violation probability $\pv(w)$ \cite{alzubaidy2016ton, schiessl2015delay}. This holds also in steady-state, i.e., in the limit $t\to\infty$. 
The tightest upper bound can be found by iterating over the parameter $\s>0$:
\begin{equation}
\pv(\wnormal) \leq \inf_{\s>0}\left\{ \Mfun{\s,\wnormal} \right\} 
\;.
\label{eq:pdelay_bound}
\end{equation}

\subsection{SNC and Scheduling}
\label{ssec:queueing_rr}
The delay analysis through stochastic network calculus as shown in Sec.~\ref{ssec:queueing} cannot be applied directly because $\Sbit(\ts)$ is zero in the time slots where the user is not scheduled, i.e., $\Sbit(\ts)$ is not i.i.d. between time slots. However, stochastic network calculus can be applied on the superframe level. The service that a user receives in superframe $i$ is denoted as $\Sbit^{(\suf)}(i)$, and is i.i.d. between superframes, because each user is scheduled exactly once per superframe of length $\suf$. The arrival process on the superframe level is given as $\Abit^{(\suf)}(i)= \alpha T$ bits, and the Mellin transform of the process $\Asnr$ in the SNR domain is $\Mellin_{\Asnr^{(T)}}(\s) = e^{\alpha T(\s-1)}$.

In case $\wtotal/\suf$ is an integer, it follows directly from \eqref{eq:pdelay_bound} that:
\begin{align}
\pv(\wtotal) & \le \inf_{\s>0}\left\{  \Mfunsuper{\s,\frac{\wtotal}{\suf}} \right\}
\label{eq:snc_pv_bound_singlegroup}
\,.
\end{align}
When the condition $\Mellin_{\Asnr^{(\suf)}}(1+\s)\Mellin_{\Ssnr^{(\suf)}}(1-\s) < 1$ holds, the kernel $\Mfunsuper{\cdot}$ converges to
\begin{align}
\Mfunsuper{\s,\frac{\wtotal}{\suf}}&= \frac{\Mellin_{\Ssnr^{(\suf)}}(1-\s)^{\frac{\wtotal}{\suf}}}{1-\Mellin_{\Asnr^{(\suf)}}(1+\s)\Mellin_{\Ssnr^{(\suf)}}(1-\s)} 
\,.
\label{eq:snc_kernel2}
\end{align}
The bound depends on the Mellin transform of the service $\Ssnr^{(\suf)}$ per superframe in the SNR-domain, which is connected to the bit-domain service process as $\Ssnr^{(\suf)}=e^{\Sbit^{(\suf)}}$. 
In the bit-domain, each user experiences a service of $\Sbit^{(\suf)}=\nd R Z$ bits per superframe, where $Z=0$ when a transmission error occurs and $Z=1$ otherwise. Therefore:
\begin{align}
\Mellin_{\Ssnr^{(\suf)}}(\s)=\expected{\left(e^{\nd \rate Z}\right)^{\s-1}}
\,.
\end{align}

In case $\wtotal/\suf$ is not an integer, some users (denoted as group 1) will be served $\lceil \wtotal/\suf\rceil$ times before the deadline, while others (group 2) will only be served $\lfloor\wtotal/\suf\rfloor$ times. For the sake of fairness, we assume that the users are assigned randomly to the slots. Then, the probability of being in the second group is $p_2=\frac{\mod(\wtotal,\suf)}{\suf}$, and $p_1 = 1-p_2$. Thus, the overall bound on the delay violation probability is given by \cite{schiessl2018misoglobecom}
\begin{align}
\pv(\wtotal) \le p_1 \Mfunsuper{\s,\left\lceil\frac{\wtotal}{\suf}\right\rceil} + p_2 \Mfunsuper{\s,\left\lfloor \frac{\wtotal}{\suf}\right\rfloor}
\;.
\label{eq:snc_pv_bound_multigroups}
\end{align}

\subsection{Delay Analysis -- Ideal Case}
\label{ssec:delay_analysis}

In the ideal case with perfect CSI and long codewords, the rate is given as $R=\log_2(1+\power\xi)$, and no errors occur ($Z=1$ with prob. $1$). In this case, we can obtain the Mellin transform of $\Ssnr^{(\suf)}$ in closed form. 
The variable $\xi$ is a central $\chi^2$ variable with $2m= 2(\Nt-K+1)$ degrees of freedom as outlined in Sec.~\ref{ssec:system_data_phy}. 
Thus, we obtain the following result:
\begin{result}
Given the transmit power $\fixpower$ and the number of scheduled users $K$, the Mellin transform of the service process is given as:
\begin{align}
\Mellin_{\Ssnr\sufss}(1-\s) &=\sum\limits_{\LL=0}^{m-1}\frac{\binom{m-1}{\LL} (-1)^{\LL}}{\Gamma(m)\fixpower^{\LL+\salt}}  e^{\frac{1}{\fixpower}}
\Gamma\left(m-\mu-\salt,\frac{1}{\fixpower}\right)
\label{eq:mellin_service_constpower_series}
\,,
\end{align}
where $m=\Nt-K+1$, $\salt\defined\frac{\s \nd}{\ln 2}$, and $\Gamma(s,x)$ is the upper incomplete Gamma function
\begin{align}
\Gamma(s,x) = \int_{x}^{\infty}t^{s-1}e^{-t}dt
\,.
\label{eq:incomplete_gamma}
\end{align}
\end{result}
\begin{proof}
Outline: Start with
\begin{align}
\Mellin_{\Ssnr\sufss}(1-\s) &=\expected{\left(e^{\nd R}\right)^{-\s}} = \int\limits_0^\infty (1+\fixpower\xi)^{-\salt} f_m(\xi)d\xi 
\,,
\end{align}
where $f_m(\xi)$ in \eqref{eq:pdf_central_chisquare_scaled} contains the term $\xi^{m-1}$, which is expanded using the binomial theorem:
\begin{align}
\xi^{m-1} &=
\frac{1}{\fixpower^{m-1}}\sum\limits_{\LL=0}^{m-1}\binom{m-1}{\LL}\left(1+\fixpower\xi\right)^{m-1-\LL}(-1)^{\LL}
\,.
\end{align} 
The result is obtained from further algebraic derivations. The details can be found in \cite{schiessl2018misoglobecom}.
\end{proof}
Thus, given the arrival rate $\alpha$ in bits per time slot and a specific choice of superframe length $\suf$, the upper bound \eqref{eq:snc_pv_bound_multigroups} on $\pv(w)$ can be obtained analytically through \eqref{eq:mellin_service_constpower_series}.

\section{Analysis -- Realistic Case}
\label{sec:analysis_icsi} 
In the previous section, we analyzed the delay performance of the multiuser MISO downlink channel and provided closed-form expressions for the Mellin transform of the service $\Ssnr\sufss$ provided to each user during a superframe of $\suf$ time slots. 
However, the analysis did not account for some of the effects that may severely deteriorate the performance of actual systems. In a real system, the transmitter first needs to acquire an estimate $\Hmmse$ of the channel matrix $\Hmat$ before computing the beamforming matrix $\Vmat$. Due to the channel estimation error, the ZFBF matrix $\Vmat$ is not perfectly matched to the actual channel $\Hmat$, and thus, the interference cannot be completely eliminated.
Furthermore, the transmitter must adapt the coding rate $\rate$ to the imperfect channel estimate $\Hmmse$.
Outages will occur whenever the rate $\rate=\rateadapt(\Hmmse)$ selected by the transmitter happens to be below the actual capacity. Moreover, when the blocklength of the channel code is small, then one cannot achieve error-free transmissions at a rate equal to the channel capacity. Instead, the transmitter must choose rates below the channel capacity in order to achieve low (but still non-zero) error probabilities. All these effects have an impact on the optimal number of scheduled users $\K$. Specifically, there are now three additional reasons to choose a small value of $\K$:
\begin{itemize}
	\item Channel estimation overhead: For each scheduled user, a dedicated training period is required. At large $\K$, this overhead severely reduces the number of symbols $\nd$ that remain for the data transmission.
	Finite blocklength effects may cause an additional performance loss when $\nd$ becomes very small.
	\item Interference: The signal for each scheduled user creates interference at the other users. A smaller number $\K$ thus reduces the interference.
	\item Backoff: In order to transmit reliably, the transmitter must often choose a rate $\rate=\rateadapt(\Hmmse)$ below the estimated capacity. 
	Reducing the number of scheduled users $\K$ increases the individual channel capacities and thus reduces the \emph{relative} impact of this backoff.
\end{itemize}
However, there is now also a major reason to increase the number of scheduled users $\K$:
\begin{itemize}
	\item Reliability: When more users are scheduled, the transmitter can schedule each user more often, which means that multiple retransmission opportunities are available for each user before the target deadline is reached. Thus, even though  scheduling many users (large $\K$) may reduce the reliability of the \emph{individual} transmissions, it may massively enhance the \emph{overall} reliability of the system with respect to the deadline.
\end{itemize}
Taking all these effects into account, our main problem remains the same: we want to determine the optimal value of $\K$ such that the overall reliability of the system with respect to the target deadline is maximized.
To solve this question, one must also solve a secondary problem: one must determine the optimal rate adaptation function $\rateadapt:\Hmmse\to\rate$. When choosing a high rate $\rate=\rateadapt(\Hmmse)$, the corresponding error probability $\perror$ will be too high. On the other hand, choosing very low rates may also lead to violations of the deadline, because then the transmitter cannot transmit all buffered data. 

In order to solve these questions, we model in the following the effects of imperfect CSI and finite blocklength channel coding.
Without loss of generality, we will consider only the signal at receiver $k=1$. The derived quantities in this section correspond to user $k=1$, which will not be indicated by a subscript. 

\subsection{Imperfect CSI}
\label{ssec:imperfect_csi_system_analysis}
We consider a time-division duplex (TDD) system, where the transmitter can estimate the channel from training sequences of length $\nt$ symbols sent by the users in the uplink. The training sequences must be mutually orthogonal, thus $\K\nt$ channel uses are required for the training of $K$ users. The SNR of the uplink channel is denoted as $\Puplink$ and known to the transmitter. 
By observing the training sequence, the transmitter can obtain the MMSE estimate $\hmmse_1$ of the channel towards user $1$.
According to \cite{caire2010multiuser}, the actual channel vector $\hvec_1$ is given in terms of the MMSE estimate $\hmmse_1$ as
\begin{equation}
\hvec_1=\hmmse_1 + \esterror_1
\,,
\end{equation}
with $\esterror_1\sim\mathcal{CN}(0,\sigmansq \eye)$ independent of $\hmmse_1$, and
\begin{equation}
\sigmansq = \frac{1}{1+\Puplink \nt}
\,.
\end{equation}
The transmitter then applies zero-forcing beamforming to create the beamforming matrix $\Vmat$ based on the estimated channel matrix $\Hmmse$. The received signal at user $1$ is given as:
\begin{align}
y_1 &= \sqrt{\power}\hvec_1^\herm \xvec + z_1
\\&=\sqrt{\power} \hvec_1^\herm \vvec_1 s_1 +\sum_{j=2}^{\K}  \sqrt{\power} \hvec_1^\herm \vvec_j s_j + z_1
\,.
\end{align}
The signal-to-interference-plus-noise ratio (SINR) is then:
\begin{align}
\sinr &= \frac{\power |\hvec_1^\herm \vvec_1|^2}{1 + \sum_{j=2}^{K} \power |\hvec_1^\herm \vvec_j|^2}
\\&= \frac{\power |(\hmmse_1 + \esterror_1)^\herm \vvec_1|^2}{1 + \sum_{j=2}^{K} \power |\esterror_1^\herm \vvec_j|^2} =\frac{\Sigpower}{1+\Intsum}
\label{eq:sinr}
\,,
\end{align}
where we have denoted the power of the signal at the receiver as $\Sigpower$ and the sum of the interference powers as $\Intsum$, and where we made use of the fact that $\hmmse_1^\herm \vvec_j=0$ for $j= 2,\ldots, K$. 
After the uplink training, the transmitter must send a downlink training sequence of $\ndl$ symbols per user, so that the receivers can also learn the channel and decode the signal \cite{caire2010multiuser}. In this work, we assume that when $\ndl\ge\nt$, the estimation error at the receiver is negligible compared to the estimation error at the transmitter.\footnote{The first reason for this assumption is that the transmitter can generally transmit at higher power than the users' devices, which may be battery-powered. Second, a small estimation error at the transmitter may lead to significant interference and to outages, whereas a small estimation error at the receiver would correspond only to a small additional noise term in the decoding process. Third, the receiver can estimate the channel not only from the dedicated training sequence, but also from the codeword itself (joint estimation and decoding) \cite{skoglund2002code}, further improving CSI at the receiver.}
For the moment, we ignore the effects of finite blocklength channel coding, and assume that data can be successfully transmitted when the data rate $\rate$ is below the instantaneous capacity $c=\log_2(1+\sinr)$. Note that for a fixed slot length of $\ntot$ symbols, only
$\nd=\ntot-K(\nt+\ndl)$
symbols remain for the data transmission.

Given the estimated channel $\Hmmse$, the transmitter must also choose a certain data rate $\rate$ for the transmission. However, due to the channel estimation error $\esterror_1$, the transmitter does not know the exact value of the SINR in \eqref{eq:sinr}. Therefore, there is a chance that the channel will be in outage, i.e., that the transmission fails. The outage probability $\pout$, conditioned on the channel estimate $\Hmmse$, is defined as
\begin{equation}
\pout = \Prob{\left.\log_2(1+\sinr)<\rate \,\right| \Hmmse}
\,.
\end{equation}
Unfortunately, analytic expressions for $\pout$ cannot be easily determined, as $\pout$ depends on the joint distribution of the received signal power $\Sigpower$ and the interference power $\Intsum$.
In order to find a solution, we consider the distributions of $\Sigpower$ and $\Intsum$ separately.
The signal-of-interest has power $\Sigpower=\power|(\hmmse_1+\esterror_1)^\herm\vvec_1|^2$. Conditioned on the known values $\hmmse_1$ and $\vvec_1$, with $\musig = \power |\hmmse_1^\herm\vvec_1|^2$ denoting the estimated receive power, $\Sigpower$ has non-central chi-square distribution with two degrees of freedom
and PDF
\begin{align}
f_{\Sigpower|\musig}(\sigpower) = \frac{1}{\power\sigmansq} e^{-\frac{\sigpower + \musig}{\power\sigmansq}} I_0\left(\frac{2\sqrt{\musig \sigpower}}{\power\sigmansq}\right)
\,.
\end{align}
The interference $\Intsum$ is the sum of $K-1$ random variables, each being exponentially distributed with mean 
\begin{align}
\expected{\power |\esterror_1^\herm \vvec_j|^2} 
&= \power \vvec_j^\herm\expected{ \esterror_1 \esterror_1^\herm }\vvec_j
= \power \sigmansq
\label{eq:variance_single_interference_term}
\,.
\end{align}
The beamforming vectors $\vvec_j$ are not mutually orthogonal, and thus the individual interference terms are correlated, so that one cannot determine the distribution of $\Intsum$, conditioned on $\Hmmse$ or on the corresponding $\Vmat$, in closed form. As $\Intsum$ is a sum of random variables, the variance of $\Intsum$ is minimal in case the individual interference terms are completely independent (all $\vvec_j$ are orthogonal), and the variance is maximal in case all the interference terms are completely correlated (all $\vvec_j$ point in the same direction). Minimum variance of the interference generally minimizes the chance that the interference $\Intsum$ is very large, and thus minimizes the outage probability $\pout$ compared to the correlated case. Similarly, we conjecture that maximum variance (due to completely correlated interferers) maximizes the outage probability $\pout$. Therefore, in the following two subsections, we use these two cases to obtain approximate lower and upper bounds on the outage probability $\pout$.\footnote{We acknowledge that considering only the variance of $\Intsum$ is not sufficient to find rigorous bounds, which would need to be based on the actual joint distribution of $\Sigpower$ and $\Intsum$, for which no analytical expressions are known. We will verify the bounds in Sec.~\ref{sec:numerics}.}

\subsubsection{Lower Bound -- Uncorrelated Interference}
When assuming that the vectors are mutually orthogonal, all the interference terms $\power |\esterror_1^\herm \vvec_j|^2$ are uncorrelated. In this case, the sum interference $\Intsum$ is given as the sum of $K-1$ independent, exponentially distributed random variables, each with mean $\power\sigmansq$.
Thus, the interference $\Intsum$ has gamma distribution with shape factor $\shape=(K-1)$ and scale $\lambda=\power \sigmansq$, whose cumulative distribution function is given as:
\begin{align}
F_\Intsum(x) \approx 1-\frac{\Gamma\left(\shape,\frac{x}{\lambda}\right)}{\Gamma(\shape)}
\,.
\end{align}
Defining $\gammao = 2^\rate-1$, the conditional outage probability can then be approximated as
\begin{align}
\pout &= \Prob{\left.\sinr < \gammao\right| \Hmmse} = \Prob{\left.\frac{\Sigpower}{1+\Intsum} < \gammao\right| \Hmmse}
\\&= \int_{0}^{\gammao}f_{\Sigpower|\musig}(\sigpower) d\sigpower + \int_{\gammao}^{\infty} \Prob{\left.\frac{\sigpower}{\gammao} - 1 < \Intsum\right| \Hmmse }f_{\Sigpower|\musig}(\sigpower)d\sigpower
\\&\approx \int_{0}^{\gammao}f_{\Sigpower|\musig}(\sigpower) d\sigpower + \int_{\gammao}^{\infty} \frac{1}{\Gamma(\shape)} \Gamma\left(\shape, \frac{\sigpower}{\lambda\gammao} - \frac{1}{\lambda}\right)f_{\Sigpower|\musig}(\sigpower)d\sigpower
\,.
\label{eq:pout_chisquare_exact}
\end{align}
In order to find a closed-form solution for the second term, we extend the results in \cite{schiessl2016imperfectcsi} to the multi-antenna scenario and derive a Gaussian approximation for $\Sigpower$:
\begin{align}
G &= \power \vvec_1^\herm(\hmmse_1+\esterror_1)(\hmmse_1+\esterror_1)^\herm\vvec_1
\\&= \underset{\musig \,\,\,\, \text{(known)}}{\underbrace{\power |\hmmse_1^\herm\vvec_1|^2}} 
+ \underset{\Sigpowergauss \,\sim\mathcal{N}(0,\sigmasigsq)}{\underbrace{2 \power |\hmmse_1^\herm\vvec_1| \real{e^{-i\angle(\hmmse_1^\herm\vvec_1)}  \esterror_1^\herm\vvec_1} }}
+ \underset{\text{negligible}}{\underbrace{\power|\esterror_1^\herm\vvec_1|^2 }}
\,.
\end{align}
We note that $\esterror_1^\herm\vvec_1$ is Gaussian and distributed as $\mathcal{CN}(0, \sigmansq)$, according to \eqref{eq:variance_single_interference_term}. The distribution of a circularly symmetric random variable is not affected by a phase shift $e^{-i\angle(\hmmse_1^\herm\vvec_1)}$, thus the second term, which is denoted as $\Sigpowergauss$, is a real-valued Gaussian variable with variance
\begin{equation}
\sigmasigsq = 4 \power^2 |\hmmse_1^\herm\vvec_1|^2 \frac{\sigmansq}{2}  = 2\sigmansq \power \musig
\,.
\end{equation}
The third term has variance $\power^2\sigman^4$, which becomes insignificant compared to $\sigmasigsq$ even for moderate training powers $\Puplink$ and training sequence lengths $\nt$. Thus, the actual receive power $\Sigpower$ can be closely approximated by $\Sigpower=\musig+\Sigpowergauss$.
Using this the Gaussian approximation for $\Sigpower$, and the finite series \cite[Eq.~(8.352.7)]{gradshteyn2007integrals} for the incomplete gamma function at integer values of $\nu$:
\begin{align}
\Gamma(\shape,x) = \Gamma(\shape) e^{-x}\sum_{m=0}^{\shape-1}\frac{x^m}{m!}
\,,
\end{align}
we find that the outage probability in case of uncorrelated interferers is approximated by:
\begin{align}
\pout &\approx
 Q\left(\frac{\musig-\gammao}{\sigmasig}\right) + \int_{\gammao}^{\infty} e^{-\frac{\sigpower}{\lambda \gammao} + \frac{1}{\lambda}} \sum_{m=0}^{\shape-1}\frac{ \left(\frac{\sigpower}{\lambda \gammao} - \frac{1}{\lambda}\right)^m}{m! \sqrt{2\pi\sigmasigsq}}  e^{-\frac{(\sigpower-\musig)^2}{2\sigmasigsq}} d\sigpower
\,.
\end{align}

We define $\mutilde = \musig -\frac{\sigmasigsq}{\gammao}$, and we obtain after some algebra, which includes the binomial expansion $\left(\frac{\sigpower}{\lambda \gammao} - \frac{1}{\lambda}\right)^m = \frac{((\sigpower -\mutilde)+(\mutilde-\gammao))^m}{(\lambda \gammao)^m}=\frac{1}{(\lambda \gammao)^m}\sum_{l=0}^{m}\binom{m}{l}(\sigpower -\mutilde)^{l}(\mutilde-\gammao)^{m-l}$:
\begin{result}
When assuming that the interferers are mutually orthogonal, the conditional outage probability for a given channel estimate $\Hmmse$ and a given rate $\rate$ can be approximated as
\begin{align}
\pout(\rate,\Hmmse) &\approx Q\left(\frac{\musig-\gammao}{\sigmasig}\right) + e^{\frac{1}{\lambda} - \frac{\musig}{\lambda\gammao} + \frac{\sigmasigsq}{2(\lambda\gammao)^2}}\sum_{m=0}^{\nu-1}\sum_{\binl=0}^{m}\frac{\binom{m}{\binl} \left(\mutilde-\gammao\right)^{m-\binl}}{(\lambda\gammao)^m m!} B_\binl(\gammao-\mutilde)
\,
\label{eq:approx_pout_gamma}
\end{align}
 with $\musig=\power|\hvec_1\vvec_1|^2$, $\gammao=2^\rate-1$ and
\begin{align}
B_\binl(x) = \int_{x}^{\infty} t^{\binl}\frac{1}{\sqrt{2\pi\sigmasigsq}} e^{-\frac{t^2}{2\sigmasigsq}} dt
\label{eq:def_B_integral}
\,.
\end{align}
\end{result}
It can be seen directly that $B_0(x)=Q\left(x/\sigmasig\right)$. Furthermore, we obtain $B_1(x)=\sqrt{\frac{\sigmasigsq}{2\pi}}e^{-\frac{x^2}{2\sigmasigsq}}$.
For values $\binl \ge 2$, integration by parts can be applied, resulting in:
\begin{align}
B_\binl(x) = \sqrt{\frac{\sigmasigsq}{2\pi}} x^{\binl-1}e^{-\frac{x^2}{2\sigmasigsq}} + (\binl-1)\sigmasigsq B_{\binl-2}(x)
\label{eq:B_integral_l2}
\,.
\end{align}
Thus, the values of $B_\binl(\gammao-\mutilde)$ for $\binl=2,\ldots,\shape-1$ can be obtained iteratively from \eqref{eq:B_integral_l2}.

\subsubsection{Upper Bound -- Correlated Interference}
In the previous subsection, we considered the case where all vectors $\vvec_j$ are orthogonal, and thus, the individual interference terms are independent. Conversely, we consider in this subsection the extreme case where the vectors $\vvec_j$ for $j=2,\ldots,K$ are identical, resulting in completely correlated interference.
This assumption results in the maximum possible variance of the interference $\Intsum$. If all $\vvec_j$ are equal, then the sum interference $\Intsum$ is equal to $(K-1)$ times the first interference term $\power|\esterror_1^\herm\vvec_2|^2$, which is exponentially distributed with mean $\power\sigmansq$. Thus, $\Intsum$ is exponentially distributed with mean $\lambdac = \power\sigmansq(K-1)$, which is equivalent to a gamma-distributed variable with shape $\shape=1$ and scale $\lambdac$. Thus, the outage probability can be approximated by \eqref{eq:approx_pout_gamma}, which for $\shape=1$ simplifies to the following result:
\begin{result}
When assuming completely correlated interference, the outage probability for a given channel estimate $\Hmmse$ and a given rate $\rate$ can be approximated as
\begin{align}
\pout(\rate,\Hmmse) &\approx Q\left(\frac{\musig-\gammao}{\sigmasig}\right) + e^{\frac{1}{\lambdac} - \frac{\musig}{\lambdac\gammao} + \frac{\sigmasigsq}{2(\lambdac\gammao)^2}} Q\left(\frac{\gammao-\mutilde}{\sigmasig}\right)
\,
\label{eq:approx_pout_exp}
\end{align}
 with $\musig=\power|\hvec_1\vvec_1|^2$ and $\gammao=2^\rate-1$.
\end{result}

\subsection{Finite Blocklength Channel Coding}
\subsubsection{Background}
When the duration of each time slot is short, the blocklength of the channel code used for the transmission is rather small. This invalidates the assumption that error-free transmissions can be achieved at a rate equal to the channel capacity $\log_2(1+\sinr)$. Instead, results for finite blocklength channel coding must be used. For AWGN (additive white Gaussian noise) channels, a well-known result is given by Polyanskiy et al. \cite[Thm.~54]{polyanskiy2010channel}, who showed that given a maximum error probability $\perror$, the achievable coding rate with $\nd$ complex channel uses at SNR $\power$ is closely approximated by
\begin{equation}
\rate_\AWGN(\nd,\perror,\power) = \log_2(1+\power)-\sqrt{\frac{\dispersion_\AWGN(\power)}{\nd}}Q^{-1}(\varepsilon)
+\mathcal{O}\left(\frac{\log n}{n}\right)
\label{eq:rate_polyanskiy}
\,,
\end{equation}
where the channel dispersion, adapted to our notation\footnote{\label{footnote_notation_fbl}We define $\Rate$ in bits instead of nats, and we have $\nd$ complex-valued channel uses, corresponding to $2\nd$ real channel uses.}, is given as \cite{polyanskiy2010channel,yang2014quasi,schiessl2016imperfectcsi}
\begin{equation}
\dispersion_\AWGN(\power) = \left(1-\frac{1}{(1+\power)^2}\right)\log_2^2(e) 
\,.
\end{equation}
However, this result for AWGN channels holds only in case the interference $\Intsum$ is zero, i.e., when $K=1$ or when the transmitter has perfect CSI and applies ZFBF. Under imperfect CSI with $K>1$, each receiver experiences interference from the signals intended for other users. In order to achieve the rate \eqref{eq:rate_polyanskiy} for user $k=1$, the transmitter would need to use a non-Gaussian codebook \cite{polyanskiy2010channel}. Thus, the other users $k=2,\ldots,K$ would be subject to non-Gaussian interference, and then \eqref{eq:rate_polyanskiy} would not hold for the other users.

Thus, we must employ different results to model the effects of finite blocklength coding. Specifically, Scarlett et al. \cite{scarlett2017dispersion} considered the performance of Gaussian codebooks under non-Gaussian noise and nearest-neighbor decoding. The authors also considered the case where $K$ sender-receiver pairs transmit concurrently, using i.i.d. Gaussian codebooks, and the receiver $k=1$ experiences i.i.d. Gaussian interference from the transmitters $k=2,\ldots,K$. These results can be directly applied to our scenario because there is no difference between interference that originates from $K$ independent transmitters and interference that originates from a single transmitter superimposing $K$ independently coded signals. Therefore, when i.i.d. Gaussian codebooks are used, a second-order approximation for the achievable coding rate is given by \cite[Eq.~(24)]{scarlett2017dispersion}
\begin{equation}
\rate_\mathrm{iid}(\nd,\perror,\power) = \log_2(1+\power)-\sqrt{\frac{\dispersion_\mathrm{iid}(\power)}{\nd}}Q^{-1}(\varepsilon) +\mathcal{O}\left(\frac{\log n}{n}\right)
\label{eq:rate_scarlett_iid}
\,,
\end{equation}
where the dispersion \cite[Eq.~(27)]{scarlett2017dispersion}, adapted to our notation\textsuperscript{\ref{footnote_notation_fbl}}, is:
\begin{equation}
\dispersion_\mathrm{iid}(\power) =  \frac{2\cdot \power}{1+\power}  \log_2^2(e) 
\label{eq:dispersion_scarlett_iid}
\,.
\end{equation}
When the transmitter picks an i.i.d. Gaussian codebook containing $2^{\nd\Rate}$ different messages, the decoding error probability at the receiver at a specific $\sinr$ can thus be approximated as
\begin{equation}
\perror(\sinr) \approx Q\left(\frac{\log_2(1+\sinr)-\rate}{\sqrt{\dispersion_\mathrm{iid}(\sinr)/\nd}}\right)
\label{eq:perror_scarlett_iid}
\,.
\end{equation}
We note that the choice of a Gaussian codebook depends only on its size, defined by the number of messages $2^{n\rate}$, and not on the exact value of $\sinr$. As a result, \eqref{eq:perror_scarlett_iid} holds even when the transmitter does not know the exact value of $\sinr$ ahead of the transmission. Thus, when the transmitter knows only the estimated channel $\Hmmse$ and chooses a coding rate $\Rate$, the overall error probability at the receiver can be approximated as\footnote{This still requires that the receivers obtain perfect CSI from the $\ndl$ downlink training symbols. However, for the single-antenna case, we showed that an expression similar to \eqref{eq:perror_avg} is accurate even when the receiver has only imperfect CSI \cite{schiessl2015delay}.}
\begin{equation}
\perror \approx \expected{\left. Q\left(\frac{\log_2(1+\sinr)-\rate}{\sqrt{\dispersion_\mathrm{iid}(\sinr)/\nd}}\right) \right|\Hmmse}
\label{eq:perror_avg}
\,,
\end{equation}
where the expectation is taken over the distribution of $\sinr$, conditioned on the estimated channel matrix $\Hmmse$. 
We note that \eqref{eq:rate_polyanskiy} and \eqref{eq:rate_scarlett_iid} are second order approximations, i.e., as $\nd\to\infty$, the term $\mathcal{O}\left(\log(n)/n\right)$ becomes insignificant compared to the second term, which decays as $\mathcal{O}\left(1/\sqrt{n}\right)$. For the AWGN channel, it was shown that the approximation \eqref{eq:rate_polyanskiy} can be accurate for blocklengths as small as $\nd\approx 200$ \cite{polyanskiy2010channel}. In our previous work \cite{schiessl2016imperfectcsi}, which considered a transmitter with only one antenna, we were able to compute a strict lower bound on the achievable coding rate, which showed that the approximation was very accurate for the considered parameters. However, we are not aware of any results that can be used to verify the accuracy of \eqref{eq:rate_scarlett_iid} and \eqref{eq:perror_avg}. 
Therefore, our results should not be seen as the actually achievable performance, but rather as approximations, which can help guide the transmitter in the difficult task of selecting the coding rate $\rate$ and the optimal number of scheduled users $K$. 

The error probability $\perror$ in \eqref{eq:perror_avg} can be obtained from Monte-Carlo simulations just like in the case of infinite blocklength.
However, in order to find an optimal rate adaptation function $\rateadapt:\Hmmse\to\rate$, the transmitter must be able to quickly determine the error probability $\perror$ for rate $\rate$ through a closed-form expression. Therefore, we apply several approximations to \eqref{eq:perror_avg} in order to obtain a closed-form expression.

\subsubsection{Closed-form Approximation}
We apply the concept of \emph{random blocklength-equivalent capacity} \cite{schiessl2016imperfectcsi}, which allows treating the effects from finite-blocklength coding in the same fashion as outages. We define the random blocklength-equivalent capacity as
\begin{equation}
C_\fblequiv = \log_2(1+\sinr)-\sqrt{\frac{\dispersion_\mathrm{iid}(\sinr)}{\nd}}\Ub
\label{eq:cap_blocklength_equiv}
\end{equation}
with $\Ub\sim\mathcal{N}(0,1)$ independent of $\sinr$. For a fixed value of $\sinr$, only $\Ub$ is random, and the blocklength-equivalent outage probability $\Prob{C_\fblequiv<\rate}$ is -- by definition -- equal to the error probability at finite blocklength $\perror(\sinr)$ given in \eqref{eq:perror_scarlett_iid}. Furthermore, it can be easily verified that when both $\sinr$ and $\Ub$ are random, $\Prob{C_\fblequiv<\rate|\Hmmse}$ is exactly equal to the expression for $\perror$ given in \eqref{eq:perror_avg}. Using this concept, we follow the further steps in \cite{schiessl2016imperfectcsi} and apply the first-order Taylor approximation \cite{schiessl2016imperfectcsi}
\begin{equation}
\log_2(x) - a \ge \log_2\left(x - \frac{x}{\log_2(e)} a\right)
\label{eq:log_taylor_approx}
\end{equation}
to $C_\fblequiv$ around $(1+\sinr)$ in order to bring $\Ub$ into the same domain as $\sinr$:
\begin{align}
C_\fblequiv &\approx \log_2\left(1+\sinr - \frac{1+\sinr}{\log_2(e)}\sqrt{\frac{\dispersion_\mathrm{iid}(\sinr)}{\nd}}\Ub\right)
\label{eq:cap_blocklength_equiv_lower}
\\&= \log_2\left(1+\frac{\Sigpower}{1+\Intsum}- \left(1+\frac{\Sigpower}{1+\Intsum}\right)\sqrt{\frac{\dispersion_\mathrm{iid}\left(\frac{\Sigpower}{1+\Intsum}\right)}{\nd\log_2^2(e)}}\Ub\right)
\label{eq:cap_blocklength_equiv_lower_step1}
\\&\approx \log_2\left(1+\frac{\musig+\Sigpowergauss}{1+\Intsum}- \left(\frac{1+\Intsum+\musig+\Sigpowergauss}{1+\Intsum}\right)
\sqrt{\frac{\dispersion_\mathrm{iid}\left(\frac{\musig+\Sigpowergauss}{1+\Intsum}\right)}{\nd\log_2^2(e)}}
\Ub\right)
\label{eq:cap_blocklength_equiv_lower_step2}
\\&\approx \log_2\left(
1+\frac{\musig+\Sigpowergauss- \left(1
	+\expected{\Intsum}
	+\musig\right)\sqrt{\frac{\dispersion_\mathrm{iid}\left(\frac{\musig}{1+\expected{\Intsum}}\right)}{\nd\log_2^2(e)}}\Ub}
{1+\Intsum}
\right)
\label{eq:cap_blocklength_equiv_lower_step3}
\\&=\log_2\left(
1+\frac{\musig + \Gcsifbl}
{1+\Intsum}
\right)
\label{eq:cap_blocklength_equiv_lower_step4}
\,.
\end{align}
In \eqref{eq:cap_blocklength_equiv_lower}, we applied the Taylor approximation. In \eqref{eq:cap_blocklength_equiv_lower_step2}, we applied the Gaussian approximation $\Sigpower\approx\musig+\Sigpowergauss$. In \eqref{eq:cap_blocklength_equiv_lower_step3}, we replaced $\Intsum$ and $\Sigpowergauss$ in the factor before $\Ub$ with their respective expectations. This is reasonable because this factor corresponds only to the variance of the term with $\Ub$. Although even small values of $\Sigpowergauss$ or $\Intsum$ can cause an outage, the same small values lead only to a small change (relative to $\musig$) in the variance of the term with $\Ub$, which does not significantly affect the distribution of $C_\fblequiv$. 
Finally, in \eqref{eq:cap_blocklength_equiv_lower_step4}, we have defined $\Gcsifbl$ as the sum of the Gaussian variable $\Sigpowergauss$ and the independent Gaussian variable $\Ub$, multiplied by a constant factor. The sum of two independent Gaussian random variables is Gaussian, and the variance of the sum is equal to the sum of the individual variances. Thus, $\Gcsifbl$ is zero-mean Gaussian with variance
\begin{equation}
\sigma^2_{\csifbl} = \sigmasigsq + \left(1+\power\sigmansq(K-1)+\musig\right)^2\frac{\dispersion_\mathrm{iid}(\frac{\musig}{1+\power\sigmansq (K-1)})}{\nd \log_2^2(e)}
\,.
\end{equation}
The error probability $\perror$ due to finite blocklength coding and imperfect CSI is given in \eqref{eq:perror_avg}, which is equal to the blocklength-equivalent outage probability $\Prob{C_\fblequiv < \rate | \Hmmse}$. Using the approximation \eqref{eq:cap_blocklength_equiv_lower_step4} for $C_\fblequiv$, we can follow the same steps as in Sec.~\ref{ssec:imperfect_csi_system_analysis} to approximate $\perror$, simply replacing $\Sigpowergauss$ by $\Gcsifbl$ in \eqref{eq:approx_pout_exp}:
\begin{result}
When assuming that the interference $\Intsum$ is completely correlated, the error probability under imperfect CSI and finite blocklength coding for a given channel estimate $\Hmmse$ and rate $\rate$ can be approximated as:
\begin{align}
\perror(\rate,\Hmmse) &\approx Q\left(\frac{\musig-\gammao}{\sigma_{\csifbl}}\right) + e^{\frac{1}{\lambdac} - \frac{\musig}{\lambdac\gammao} + \frac{\sigma_{\csifbl}^2}{2(\lambdac\gammao)^2}} Q\left(\frac{\gammao-\mutilde}{\sigma_{\csifbl}}\right)
\label{eq:approx_perror_exp}
\end{align}
with $\musig=\power|\hvec_1\vvec_1|^2$, $\gammao=2^\rate-1$, $\lambdac = \power\sigmansq(K-1)$, and $\mutilde = \musig -\frac{\sigma_{\csifbl}}{\gammao}$.
\end{result}
Our numerical evaluations show that \eqref{eq:approx_perror_exp} is an upper bound to $\perror$ in \eqref{eq:perror_avg} because we assumed correlated interference and also because the Taylor approximation \eqref{eq:log_taylor_approx} was designed to obtain a lower bound on $C_\fblequiv$. 
In case the interference terms are uncorrelated, the error probability $\perror$ under finite blocklength coding can be obtained by replacing $\sigmasig$ with $\sigma_{\csifbl}$ in \eqref{eq:approx_pout_gamma} and \eqref{eq:def_B_integral}. However, due to the Taylor approximation \eqref{eq:log_taylor_approx}, the resulting expression is no longer a lower bound to $\perror$. Therefore, we use only \eqref{eq:approx_perror_exp} to estimate the effects of finite length coding.

\subsection{Delay Analysis}
\label{ssec:delay_analysis_icsi}
For the scenario with perfect CSI, we presented closed-form expressions for the Mellin transform of the SNR-domain service process $\Ssnr\sufss=e^{\nd\Rate}$ in Sec.~\ref{ssec:delay_analysis}. These closed-form expressions can be used to compute the kernel function $\Mfunsuper{\s,w/\suf}$ in \eqref{eq:snc_kernel2}, which is an upper bound on the delay violation probability $\pv(w)$. 

However, in case of imperfect CSI, finding an upper bound on $\pv(w)$ is a much more difficult task, because it is still unclear what rate $\Rate=\rateadapt(\Hmmse)$ the transmitter should choose. 
The optimal rate adaptation function $\rateadaptopt$ must find a balance between the rate $\rate=\rateadapt(\Hmmse)$ and the corresponding error probability $\perror(\rateadapt(\Hmmse),\Hmmse)$ such that the reliability of the system with respect to the deadline $w$ is maximized. Specifically, we want to find
\begin{align}
\rateadaptopt = \argmin_{\rateadapt} \quad \pv(w)
\,.
\end{align}
Because $\pv(w)$ cannot be determined in analytical form, we follow our previous work \cite{schiessl2016imperfectcsi} and perform the optimization based on the analytical upper bound $\Mfunsuper{\s,w/\suf}$ on $\pv(w)$. First, we fix a parameter $\s>0$ and determine
\begin{align}
\rateadaptopt_\s &= \argmin_{\rateadapt} \quad\Mfunsuper{\s,w/\suf,\rateadapt}
\label{eq:rateadapt_problem_specific_s}
\\&=\argmin_{\rateadapt} \quad\Mellin_{\Ssnr\sufss}(1-\s)
\label{eq:rateadapt_problem_s_mellin}
\\&=\argmin_{\rateadapt} \quad\expected{\left(1-\perror(\rateadapt(\Hmmse),\Hmmse)\right)e^{-\s \nd\rateadapt(\Hmmse)}+\perror(\rateadapt(\Hmmse),\Hmmse)}
\label{eq:rateadapt_problem_s_expected}
\,,
\end{align}
where in \eqref{eq:rateadapt_problem_specific_s}, we specifically denoted that the kernel $\Mfunsuper{\cdot}$ depends implicitly on the rate adaptation function $\rateadapt$. The second step \eqref{eq:rateadapt_problem_s_mellin} follows directly by inspecting \eqref{eq:snc_kernel2}. Then, the optimal rate adaptation function $\rateadaptopt$ can be found by iterating over $\s>0$:
\begin{align}
\rateadaptopt &\approx \argmin_{\rateadaptopt_\s} \quad\inf_{\s>0}\quad \Mfunsuper{\s,w/\suf,\rateadaptopt_\s}
\label{eq:rateadapt_problem}
\,.
\end{align}

In order to solve this problem, we use the closed-form approximation for the error probability derived in the previous section, which depends on $\Hmmse$ only through the estimated SNR $\musig=\power|\hvec_1\vvec_1|^2$. Thus, we need to take the expected value in \eqref{eq:rateadapt_problem_s_expected} only with respect to the distribution of $\musig$, which has $\chi^2$ distribution with $2m$ degrees of freedom. We quantize this distribution to points $\musig_i$ with $i=1,\ldots,\Nsnrpoints$, where the probability of each value is denoted as $p_\musig(i)$.
For each quantized value $\musig_i$, we can choose a rate $\Rate_i$ and obtain $\perror(\rate_i,\musig_i)$ from
\eqref{eq:approx_perror_exp}. Therefore,
\begin{align}
\Mellin_{\Ssnr\sufss}(1-\s) \approx \sum_i p_\musig(i) \left((1-\perror(\rate_i,\musig_i))e^{-\s\nd\rate_{i}} + \perror(\rate_i,\musig_i)\right)
\label{eq:mellin_quantized}
\,.
\end{align}
In order to find the optimal rate adaptation function $\rateadaptopt_\s$, we need to determine the optimal rates $\rate^*_i$ that minimize \eqref{eq:mellin_quantized}. This can be achieved by generating a vector of quantized rates $\rate_{j}$ with $j=1,\ldots,\Nrates$, and finding
\begin{align}
\rate^*_i = \argmin_{\rate_{j}} \quad (1-\perror(\rate_{j},\musig_i))e^{-\s\nd\rate_{j}} + \perror(\rate_{j},\musig_i))
\,.
\end{align}
The optimal rate adaptation function $\rateadaptopt$ can then be obtained by repeating this process for different values of $\s>0$. This can be done efficiently, as the error probabilities $\perror(\rate_{j},\musig_i)$ do not depend on $\s$.

\section{Numerical Results}
\label{sec:numerics}
For the numerical evaluations of our results, we first investigate in Sec.~\ref{ssec:numerics_pcsi} the channel hardening effect under the basic, ideal system model. Then, in Sec.~\ref{ssec:numerics_icsi_validate} we validate the bounds on the outage probability under imperfect CSI. In Sec.~\ref{ssec:numerics_icsi} we investigate how imperfect CSI affects the delay performance. In Sec.~\ref{ssec:numerics_fbl}, we also take the effects of finite blocklength channel coding into account, and confirm by Monte Carlo simulations that the analytical bounds on the error and outage probabilities lead to upper bounds on the delay violation probability.

\subsection{Ideal case}
\label{ssec:numerics_pcsi}
\begin{figure}[t!]
	\centering
	\vspace{2mm}
	\subfloat[\label{subfig:m8_rate}]{%
		\includegraphics[width=0.93\figurewidth]{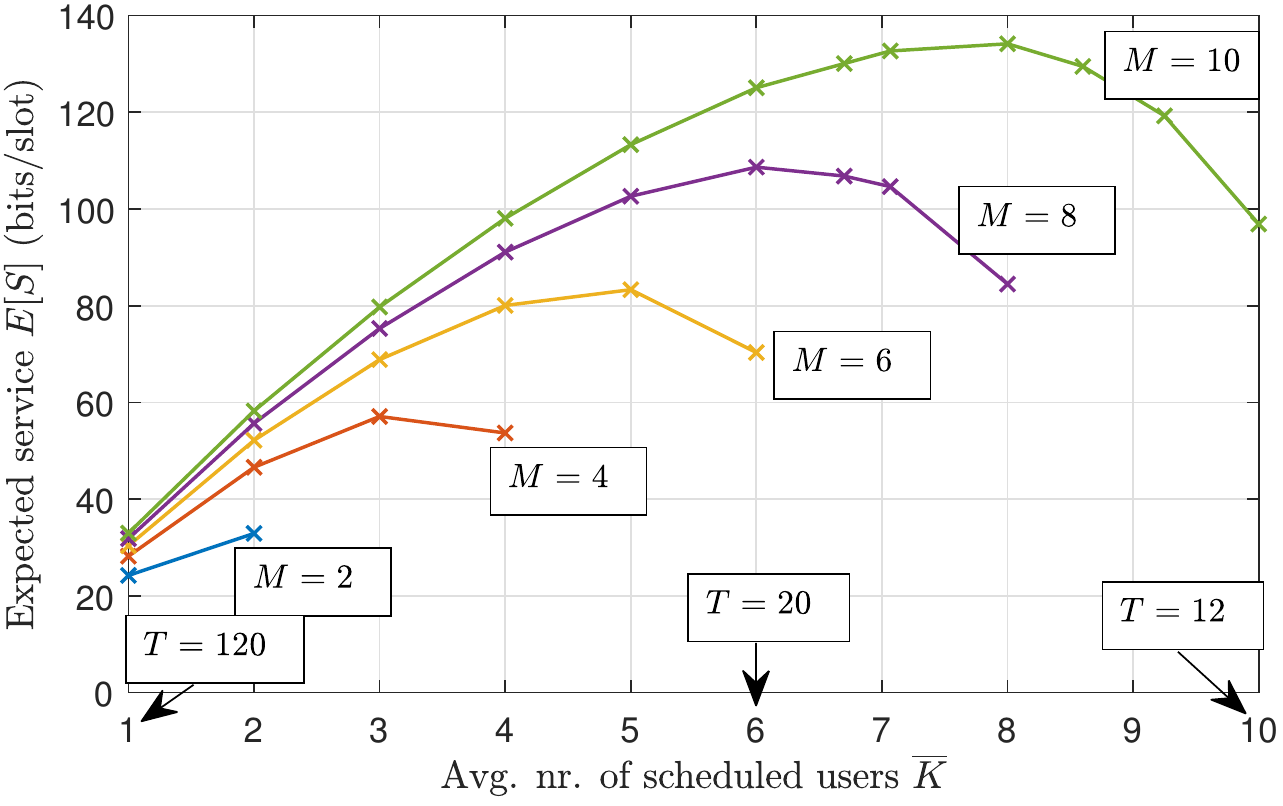}
	}
	\ifthenelse{\equal{\twocols}{true}}{\vfill}{~}
	\subfloat[\label{subfig:m8_zfbf}]{%
		\includegraphics[width=0.93\figurewidth]{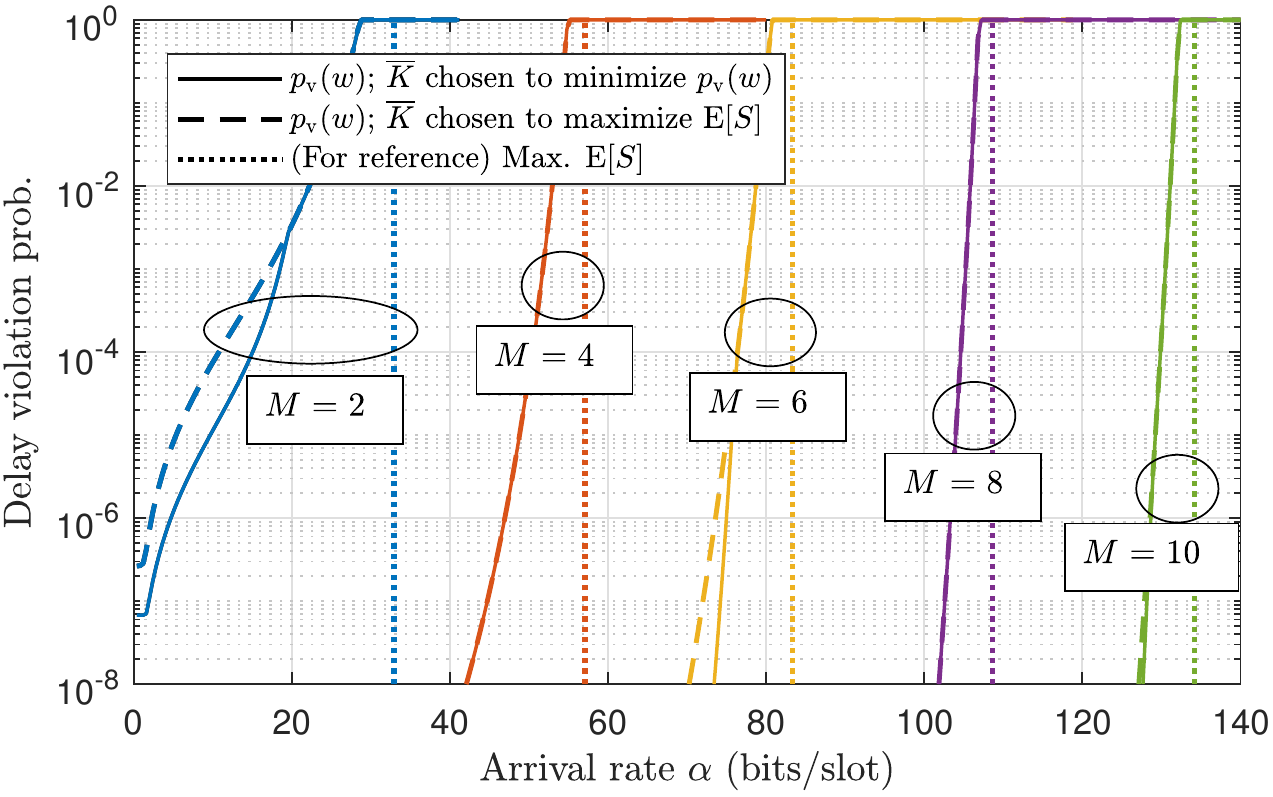}
	}
	\vfill
	\caption{ $\Ktot=120$ users, $\nd=400$ symbols, $\Ptot=20~\mathrm{dB}$. (a) Expected service $\expected{\Sbit}$ in bits per slot for $\Nt\in\{2,4,6,8,10\}$. (b) Bound on the delay violation probability $\pv(w)$ for $w=120$ slots vs. arrival rate $\alpha$, compared to expected service.}
	\label{fig:results_expservice}
\end{figure}

For the ideal scenario with perfect CSI, we first show in Fig.~\ref{fig:results_expservice} the expected service rate $\expected{S}$ of the system per time slot
versus the number of scheduled users $\Kavg$, for different numbers of transmit antennas $\Nt$. The expected service per time slot is given as
\begin{equation}
\expected{S} = \frac{1}{\suf}\expected{S^{(\suf)}} =\frac{1}{\suf}\expected{\nd R Z}
\,.
\end{equation}
In the ideal case, no errors occur, i.e., we always have $Z=1$.
The total number of users is $\Ktot=120$. The transmitter can then choose for example a superframe length of $\suf=40$ slots and schedule $\Kavg=3$ users in each time slot.
We observe that the expected service $\expected{S}$ first increases in $\Kavg$ and then decreases. This is not surprising. For example, increasing $\Kavg$ from 1 to 2 doubles the multiplexing gain, but barely affects the beamforming gain. Contrary to that, at large $\Kavg$, an increase in $\Kavg$ leads only to a minor increase in multiplexing gain, but a large decrease in the beamforming gain.

In Fig.~\ref{subfig:m8_zfbf}, we show the delay violation probability $\pv(w)$ vs. the arrival rate $\alpha$. For each data point, we select the number of scheduled users $\Kavg$ such that $\pv(w)$ is minimized. However, we note that there are only few cases where the optimization over $\Kavg$ improves the performance compared to the value $\Kavg$ that maximizes the expected service rate $\expected{\Sbit}$.
For $\Nt\in\{6,8,10\}$, we observe that $\pv(w)< 10^{-8} $ even when the arrival rate $\alpha$ is only 10\% below the expected service rate $\expected{\Sbit}$. 
Then, $\pv(w)$ rises sharply to 1 as $\alpha$ increases. 
Thus, we observe significant channel hardening and a deterministic zero/one queueing behavior for the considered parameters, even with a fairly small number of antennas. In the following, we will investigate whether this observation holds also when we consider a more realistic system model.

\subsection{Imperfect CSI -- Validation}
\label{ssec:numerics_icsi_validate}
\begin{figure}[t!]
	\centering
	\vspace{2mm}
	\subfloat[\label{subfig:m8_rate_x_pout_k5}]{%
		\includegraphics[width=0.93\figurewidth]{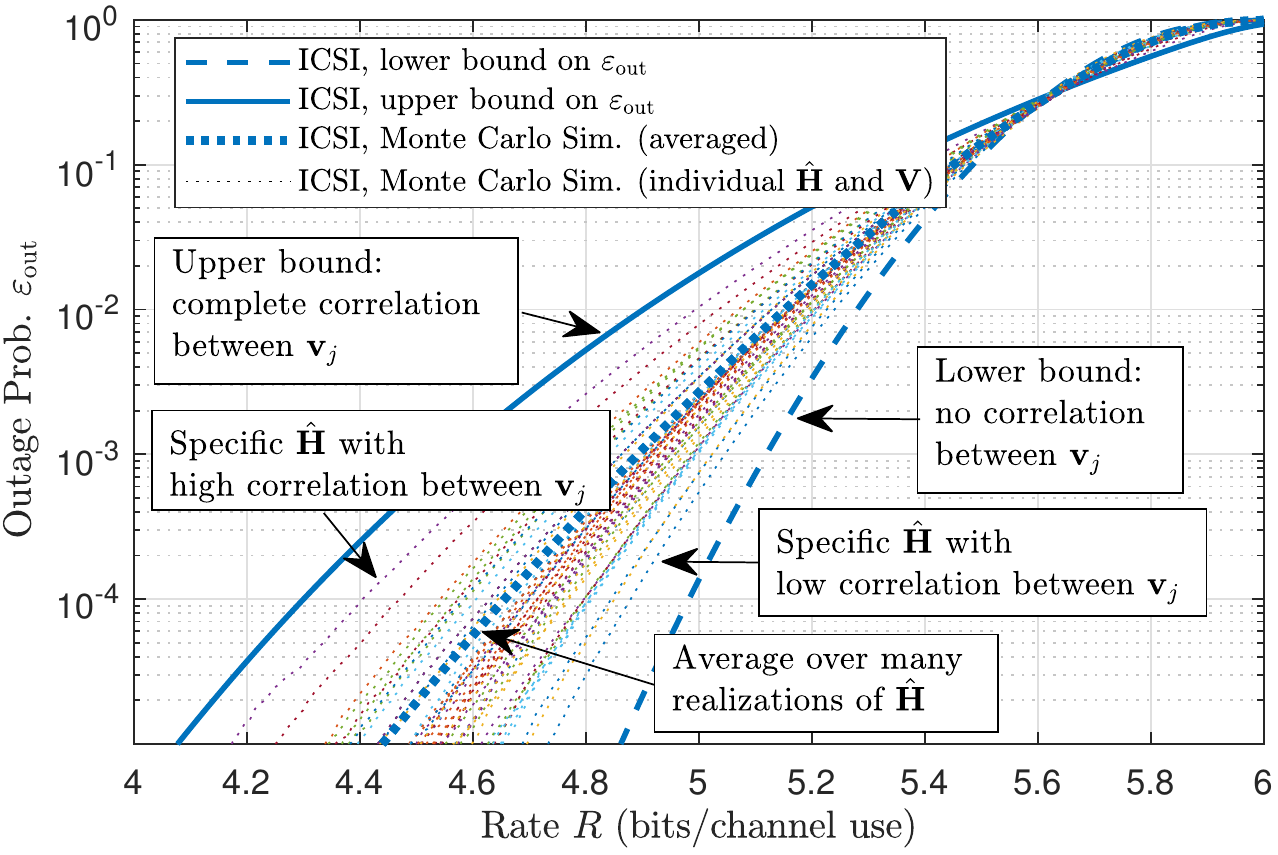}
	}
\ifthenelse{\equal{\twocols}{true}}{\vfill}{~}
	\subfloat[\label{subfig:m8_rate_x_pout_k2}]{%
		\includegraphics[width=0.93\figurewidth]{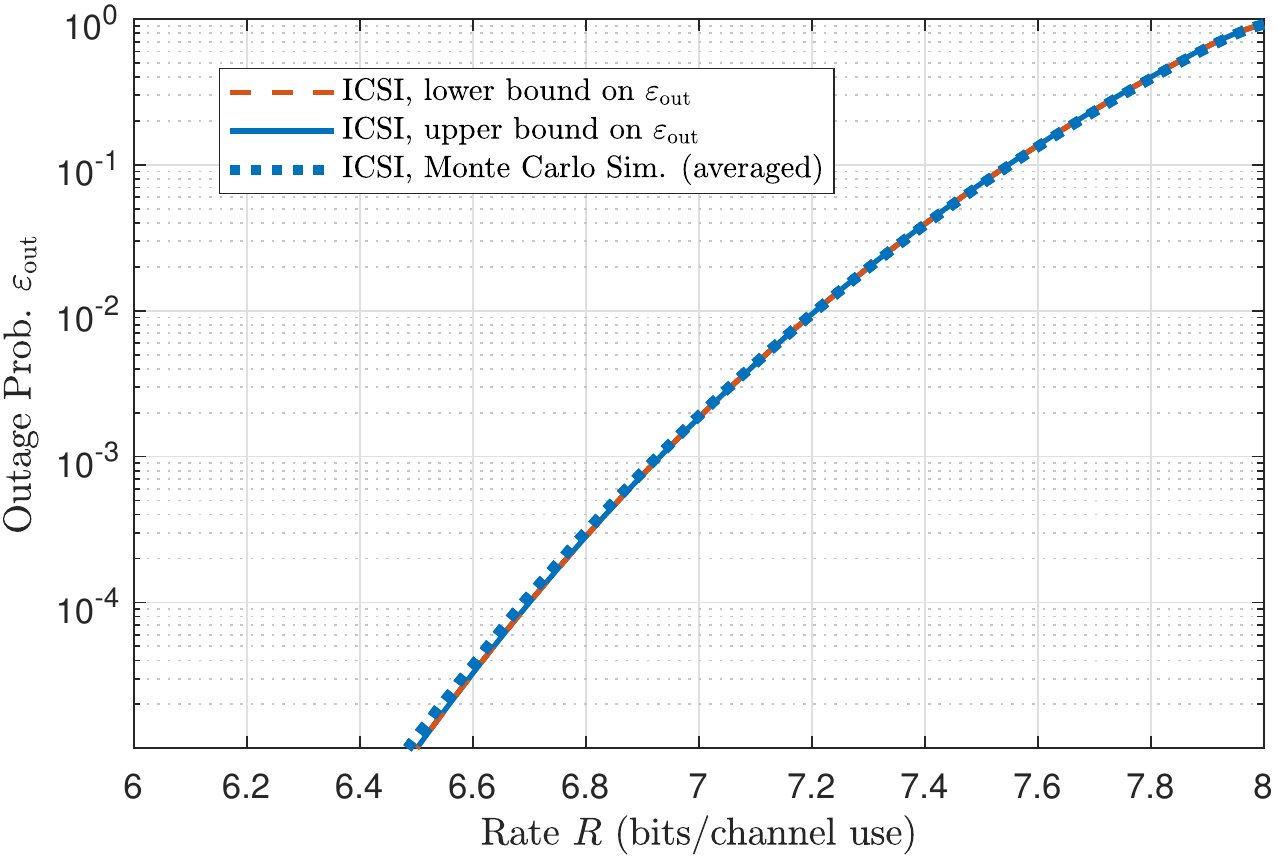}
	}
	\vfill
	\caption{Validation of lower and upper bounds on $\pout$ using Monte Carlo simulations: $\pout$ vs. $\rate$ for $\Nt=8$, $\Ptot=20~\mathrm{dB}$, $\Puplink=15~\mathrm{dB}$, $\nt=10$. (a) $K=5$, $\log_2(1+\musig)\approx 6$ bits (b) $K=2$, $\log_2(1+\musig)\approx 8$ bits.}
	\label{fig:rate_x_pout}
\end{figure}

In Fig.~\ref{fig:rate_x_pout}, we validate the lower bound \eqref{eq:approx_pout_gamma} and the upper bound \eqref{eq:approx_pout_exp} on the outage probability $\pout$. 
First, in Fig.~\ref{subfig:m8_rate_x_pout_k5} we consider $\Nt=8$ antennas with $K=5$ users. We investigate the outage probability vs. the data rate when the estimated capacity $\log_2(1+\musig)$ of the channel is 6 bits/channel use (this is close to the mean value). For the simulations, we generate matrices $\Hmmse$ and compute the corresponding beamformers $\vvec_j$ until we find a matrix $\Hmmse$ such that $\log_2(1+\musig)=\log_2(1+\power|\hmmse_1\vvec_1|^2)$ is between 5.99 and 6.01 bits. Then, we generate at least $10^6$ instances of the estimation error $\esterror_1$, in order to obtain the distribution of the actual $\sinr$, conditioned on the channel estimate $\Hmmse$. From this, the outage probability $\pout$ at rate $\rate$, conditioned on a specific $\Hmmse$, can be obtained empirically. 

Fig.~\ref{subfig:m8_rate_x_pout_k5} shows the resulting $\pout$ for several instances of $\Hmmse$, as well as $\pout$ averaged over 1000 instances of $\Hmmse$.
The average over $\pout$ (bold dotted curve) has the following meaning: when the transmitter estimates the channel capacity to be $\log_2(1+\musig)\approx 6 $ bits/channel use, the transmitter must choose a rate of $\rate\approx 4.6$ bits to achieve $\pout<10^{-4}$.
However, repeating these Monte Carlo simulations for many different values of the estimated SNR $\musig$ is computationally prohibitive. 
Instead, we want to determine the rate $\rate$ from the analytical bounds. The analytical upper bound on $\pout$ is best suited when high reliability is desired: in order to keep $\pout$ below a target of e.g. $10^{-4}$, the transmitter can choose a rate $\rate\approx 4.3$ bits/channel use. This is a robust choice, as the actual $\pout$ will be below the target of $10^{-4}$. The lower bound on $\pout$ allows bounding the performance from above: When choosing a rate $\rate\approx 5$ bits, then $\pout$ will be higher than $10^{-4}$. 
Fig.~\ref{subfig:m8_rate_x_pout_k5} shows also
that the correlation between the beamforming vectors $\vvec_j$ seems to be quite low for many instances of $\Hmmse$. Despite that, the rare instances of $\Hmmse$ where the correlation is high seem to have a strong impact on $\pout$.

In Fig.~\ref{subfig:m8_rate_x_pout_k2}, we consider $K=2$ users and channels $\Hmmse$ where $\log_2(1+\musig)$ is between 7.99 and 8.01 bits. For $K=2$, there is only a single interferer and therefore no difference between the lower bound \eqref{eq:approx_pout_gamma} and the upper bound \eqref{eq:approx_pout_exp}. The Monte Carlo simulations match the analytical results almost exactly. We note that for $K=2$, the correlation between the signal power $\Sigpower$ and the interference $\Intsum$ can sometimes lead to a tiny increase in $\pout$. Therefore, \eqref{eq:approx_pout_exp} is not a strict upper bound on $\pout$. However, in all considered scenarios, the differences were negligible.

\subsection{Imperfect CSI -- Results}
\label{ssec:numerics_icsi}
\begin{figure}[t!]
	\centering
	\vspace{2mm}
	\subfloat[\label{subfig:m8_rate_icsi}]{%
		\includegraphics[width=0.93\figurewidth]{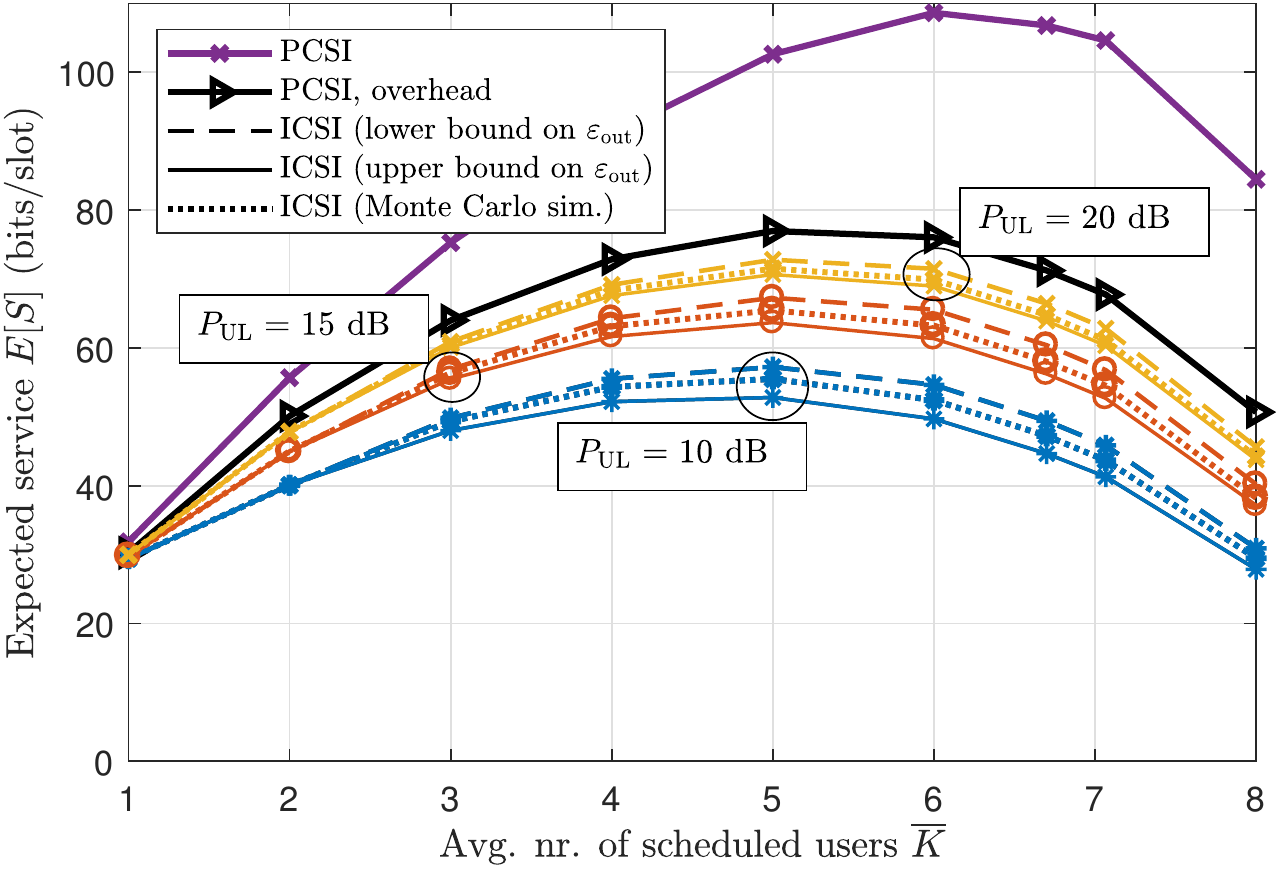}
	}
	\ifthenelse{\equal{\twocols}{true}}{\vfill}{~}
	\subfloat[\label{subfig:m8_arrivals_x_pv}]{%
		\includegraphics[width=0.93\figurewidth]{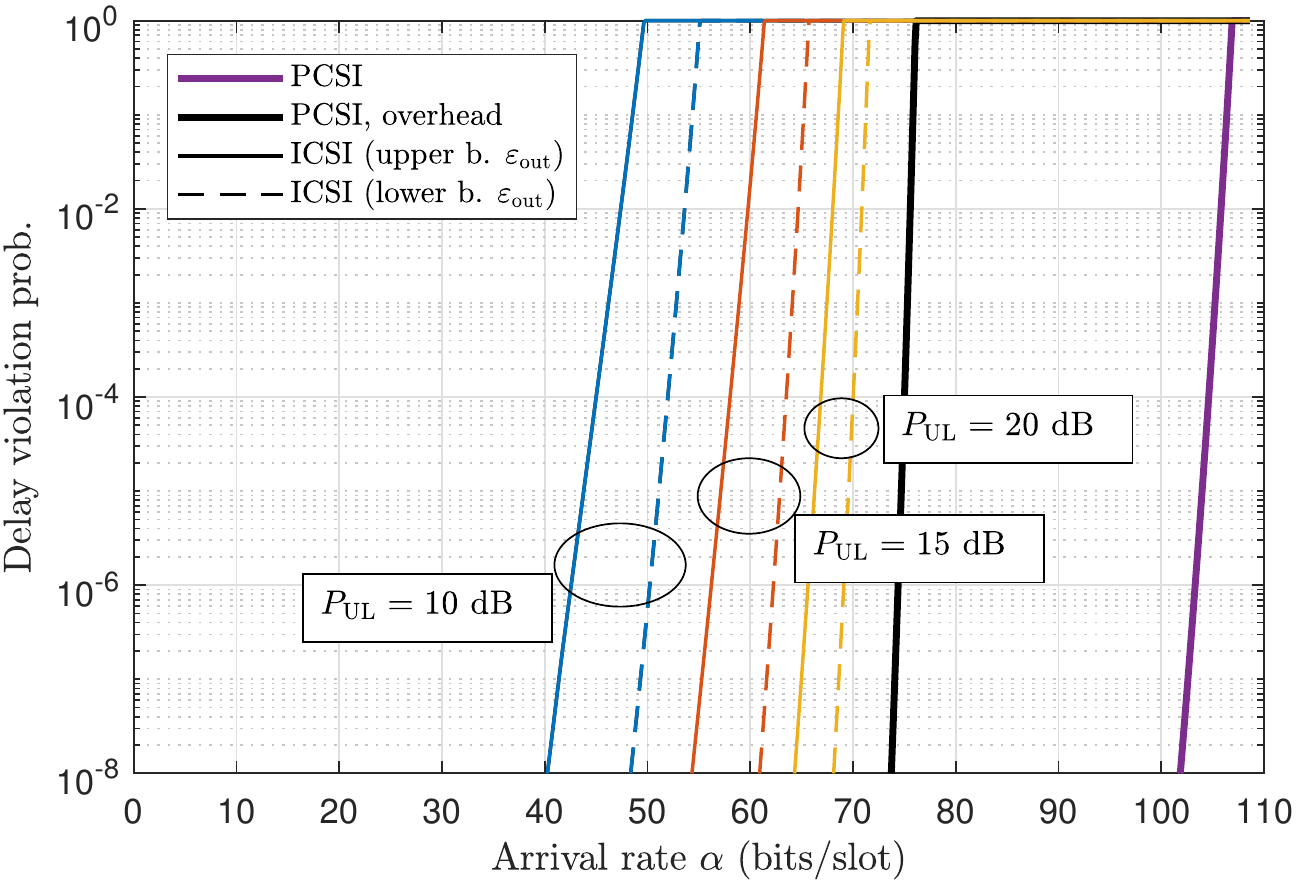}
	}
	\vfill
	\caption{ $\Ktot=120$ users, $\Nt=8$, $\ntot=400$ symbols, $\Ptot=20~\mathrm{dB}$, different values of $\Puplink$. (a) Expected service $\expected{\Sbit}$ in bits per slot vs. $\Kavg$. (b) Bound on the delay violation probability $\pv(w)$ vs. $\alpha$. Deadline $w=120$ slots.}
	\label{fig:results_icsi1}
\end{figure}

In Fig.~\ref{fig:results_icsi1}, we study the how imperfect CSI affects the performance. First, in Fig.~\ref{subfig:m8_rate_icsi} we show the expected service $\expected{\Sbit}$ per slot vs. the number of scheduled users $\Kavg$ for $\Nt=8$. The curve for perfect CSI (PCSI), while assuming no overhead, was already shown in Fig.~\ref{subfig:m8_rate} in the previous section. We now find that the performance massively deteriorates when considering channel estimation and imperfect CSI. First of all, the thick black curve shows results where CSI is still assumed to be perfect, but an overhead of $\nt=10$ and $\ndl=10$ symbols for the uplink and downlink training is taken into account. The overhead already leads to a massive performance loss at large $\Kavg$. Now, we consider three values of $\Puplink\in\{10,15,20\}~\mathrm{dB}$ to show different levels of CSI quality.
In all cases, we plot three different performance curves, corresponding to the lower bound on $\pout$ in \eqref{eq:approx_pout_gamma} (dashed curve), to the upper bound on $\pout$ in \eqref{eq:approx_pout_exp} (thin solid curve), and to Monte Carlo simulations (dotted curve).\footnote{For the Monte Carlo simulations, we use the rate adaptation function $\rateadapt$ that was obtained for the upper bound (solid curve). As expected, the simulations show better performance, because the actual $\pout$ is below the upper bound.}
We find that the two bounds for the outage probability $\pout$ match fairly well, and that the results for the Monte Carlo simulations always lie between the bounds. 
Overall, we find that imperfect CSI has quite a strong impact on the system performance when $\Kavg$ is large, but almost no impact when $\Kavg=1$. Surprisingly, this does not lead to a large change in the optimal value of $\Kavg$, which reduces only from 6 (perfect CSI) to 5 (imperfect CSI).
The results show that the optimal number of scheduled users $\Kavg$ is always correctly predicted by the bounds.

Finally, we investigate in Fig.~\ref{subfig:m8_arrivals_x_pv} whether the zero/one behavior with respect to the delay violation probability $\pv(w)$ is maintained when considering imperfect CSI. Despite the possibility outages, and the backoff that is required to achieve low outage probabilities, we find that the zero/one behavior of the system remains (even though the slopes become less steep): The delay violation probability is close to zero when the arrival rate $\alpha$ is around 20\% below the expected service $\expected{\Sbit}$, and is equal to one when $\alpha>\expected{\Sbit}$. These results stand in stark contrast to the findings for the single-antenna case \cite{schiessl2016imperfectcsi}, where imperfect CSI and finite blocklength had only a moderate impact on the expected service, but a dramatic impact on the performance under delay constraints.

\subsection{Finite Blocklength Coding}
\label{ssec:numerics_fbl}
\begin{figure}[t!]
	\centering
	\vspace{2mm}
	\subfloat[\label{subfig:m8_rate_x_eps}]{%
		\includegraphics[width=0.93\figurewidth]{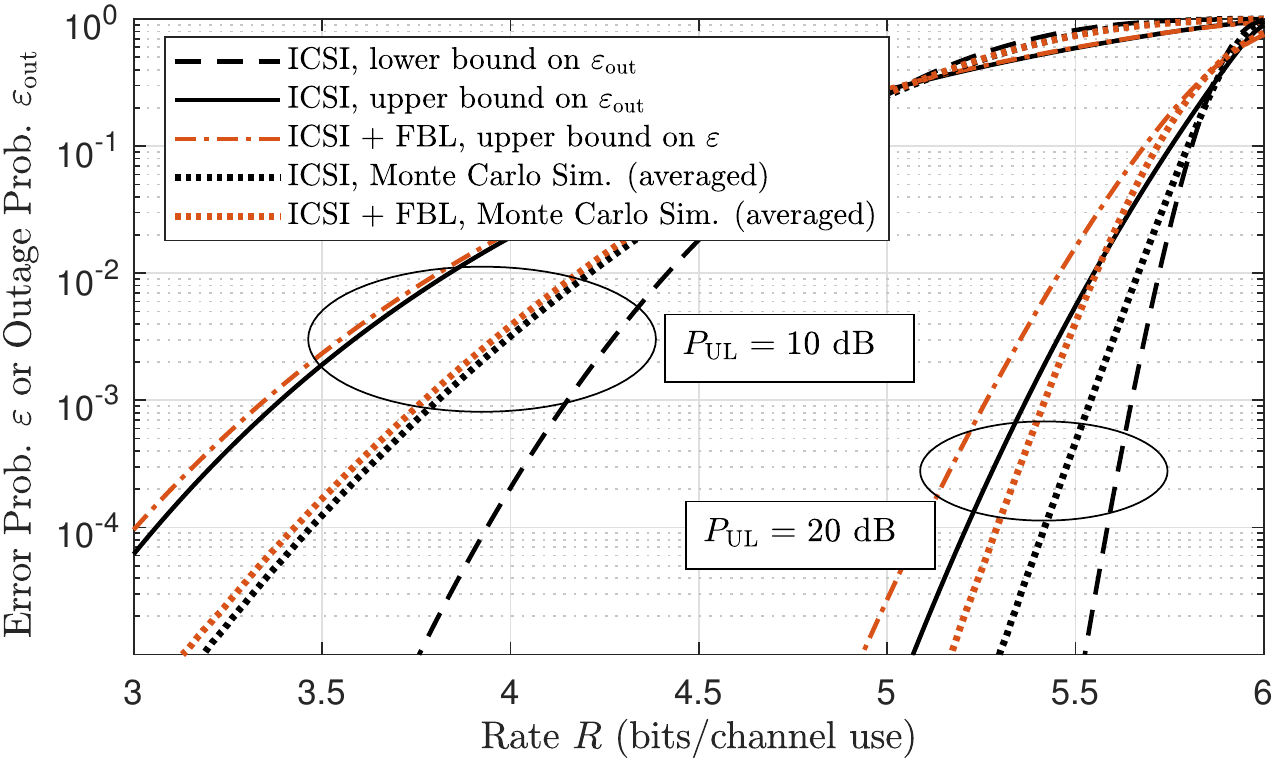}
	}
	\ifthenelse{\equal{\twocols}{true}}{\vfill}{~}
	\subfloat[\label{subfig:m8_arrivals_x_pv_fbl}]{%
		\includegraphics[width=0.93\figurewidth]{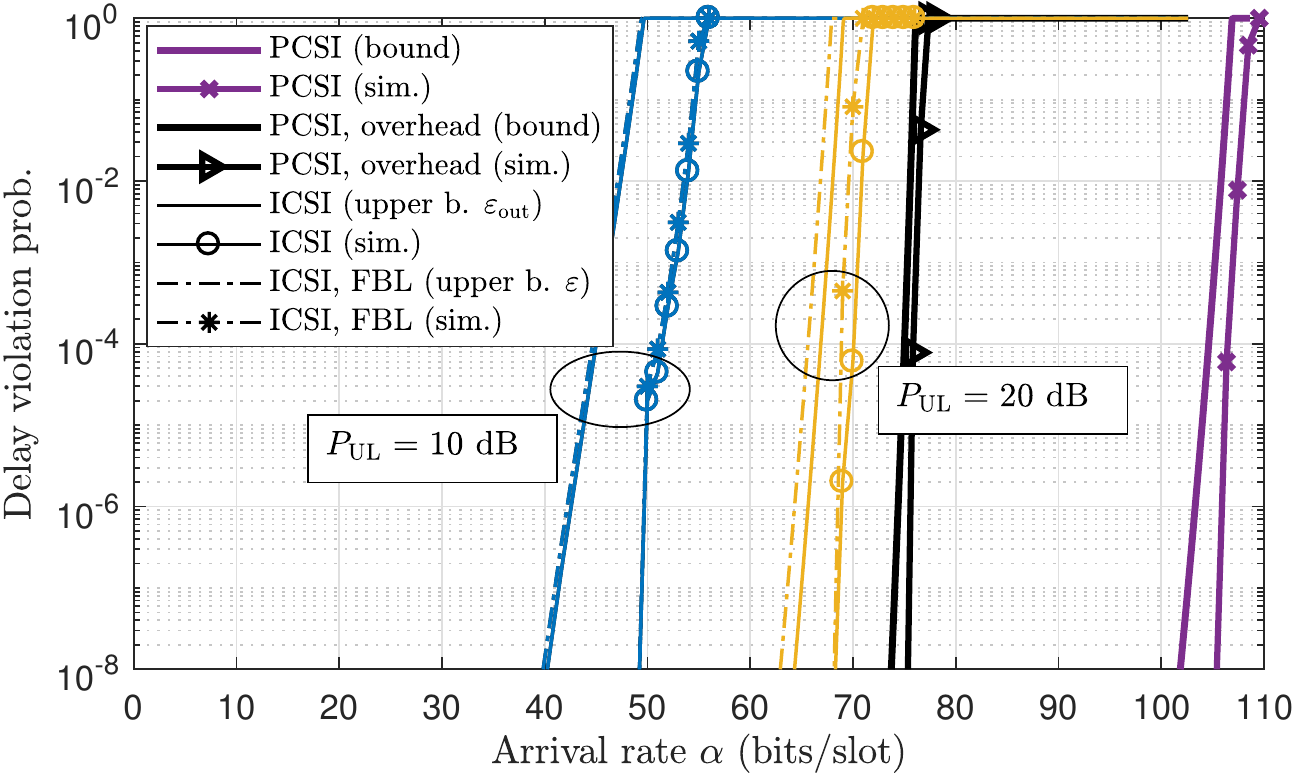}
	}
	\vfill
	\caption{ $\Ktot=120$ users, $\Nt=8$, $\ntot=400$ symbols, $\Ptot=20~\mathrm{dB}$, different values of $\Puplink$. (a)  Error/outage probabilities vs. rate $\rate$ when $\log_2(1+\musig)\approx 6$, $\K=5$. (b) Delay violation probability (both analytical bounds and Monte Carlo simulations over $10^7$ time slots) vs. arrival rate $\alpha$.}
	\label{fig:results_icsi_fbl}
\end{figure}

In Fig.~\ref{subfig:m8_rate_x_eps}, we investigate the impact of finite blocklength coding. For the simulations, we use the same methods that were used for Fig.~\ref{subfig:m8_rate_x_pout_k5}, which results in random samples of $\sinr$. The error probability $\perror$ at finite blocklength \eqref{eq:perror_avg} can be obtained by computing $\perror(\sinr)$ in \eqref{eq:perror_scarlett_iid} for each realization of $\sinr$, and then taking the average.
We find that finite length coding has very little impact on the performance when the CSI quality is poor ($\Puplink=10~\mathrm{dB}$). 
However, when the quality of the channel estimates increases, finite blocklength effects are more pronounced. Nevertheless, even at $\Puplink=20~\mathrm{dB}$, the system loses only 0.1 bits in the rate due to finite blocklength effects. Although this performance loss is small, it cannot be ignored when high reliability is desired. We note that the upper bounds on $\pout$ and on $\perror$ correctly predict the performance loss of 0.1 bits that can be observed in the simulations. 
Therefore, we use in the following only the upper bounds.

In Fig.~\ref{subfig:m8_arrivals_x_pv_fbl}, we investigate the delay performance under finite length coding. 
First, we consider only the analytical bounds. At $\Puplink=10~\mathrm{dB}$, the performance loss is dominated by imperfect CSI, and finite blocklength effects are negligible. When the accuracy of the CSI increases ($\Puplink=20~\mathrm{dB}$), finite blocklength effects cause a small performance penalty. 
However, the delay violation probability maintains its zero/one behavior.
We conclude that finite blocklength effects often have a much smaller performance impact than imperfect CSI. This is in line with previous results for the single-antenna case \cite{schiessl2016imperfectcsi}.

Finally, we consider also the Monte Carlo simulations of the queueing system. We can confirm that the actual delay violation probability $\pv(w)$ observed in the simulations is always below the analytical bounds. However, there is a gap between the upper bound and the simulation results. This is mostly because the upper bounds on $\pout$ and $\perror$ are not perfectly tight.
Nevertheless, the bounds are useful, as they predict quite accurately that a system with $\Puplink=10~\mathrm{dB}$ can only support an arrival rate $\alpha$ of 40 to 50 bits per slot, which is much less than the roughly 100 bits per slot in the ideal (perfect CSI) model.
Additionally, we note the analytical bounds correctly predict that finite blocklength coding causes only a small performance loss for the considered scenario.

\section{Conclusions}
\label{sec:conclusions}
We considered the delay performance of the multiuser MISO downlink under ideal and under realistic assumptions. Under ideal assumptions, multiple antennas create an almost deterministic queueing behavior, i.e., the system can achieve very high reliability with respect to the deadline, as long as the average transmission rate is large enough.
When considering imperfect CSI and finite blocklength coding, we observed a massive degradation of the average transmission rate. Nevertheless, we found that the system maintains the same qualitative behavior: when the average transmission rate exceeds the arrival rate, the system can still achieve very high reliability.
While it has long been known that multi-antenna technology can greatly improve the reliability, it is still surprising that the system remains extremely reliable even under non-ideal assumptions.


\appendices
\section{Analysis for non-integer $\Kavg$}
\label{appendix_groupsAB}
The average number of scheduled users per slot is defined as $\Kavg\defined\Ktot/ \suf$. In case $\Kavg$ is not an integer number, the scheduler must sometimes select more than $\Kavg$ users, sometimes less.
Specifically, the transmitter schedules $K_\aslot=\left\lceil\, \Kavg\, \right\rceil$ in $\suf_\aslot$ time slots, and $K_\bslot=\lfloor \Kavg \rfloor$ in $\suf_\bslot=\suf-\suf_\aslot$ time slots, where $\suf_\aslot$ and $\suf_\bslot$ can be determined from $\suf_\aslot K_\aslot+\suf_\bslot K_\bslot=\Ktot$. 
As the power $\Ptot$ is still shared equally between the $K_\aslot$ or $K_\bslot$ scheduled users, each users codeword is transmitted with power $\fixpower_{\aslot}=\Ptot/K_{\aslot}$ or $\fixpower_{\bslot}=\Ptot/K_{\bslot}$. Thus, the users in the type $\bslot$ slots benefit both from a higher beamforming gain and from a higher transmit power. 
To maintain fairness between all users, the users are randomly assigned to the $\aslot$ or $\bslot$ slots in each superframe.
The probability of being scheduled in a type $\aslot$ or type $\bslot$ slot is given as $p_{\aslot}=K_{\aslot} \suf_{\aslot}/\Ktot$ and $p_{\bslot}=K_{\bslot} \suf_{\bslot}/\Ktot$, respectively \cite{schiessl2018misoglobecom}. 

The Mellin transform of the service $\Ssnr\sufss$ experienced by each user can be obtained by averaging over the Mellin transforms of the service process for the type $\aslot$ or $\bslot$ slots:
\begin{equation}
\Mellin_{\Ssnr\sufss}(1-\s) = p_{\aslot}\Mellin_{\Ssnr\sufss|K_\aslot}(1-\s) + p_{\bslot}\Mellin_{\Ssnr\sufss|K_\bslot}(1-\s)
\label{eq:mellin_service_fixedpowers}
\,,
\end{equation}
where the Mellin transform $\Mellin_{\Ssnr\sufss|K}(\s)$ for a specific number of users $K$ is given by \eqref{eq:mellin_service_constpower_series} for perfect CSI, or by \eqref{eq:mellin_quantized} when considering imperfect CSI and finite blocklength effects.

\bibliographystyle{IEEEtran}
\IfFileExists{../../PHD2.bib}
{\bibliography{../../PHD2}}
{
	\IfFileExists{../PHD2.bib}
	{\bibliography{../PHD2}}
	{\bibliography{multiuser_mimo_journal}}
}

\end{document}